\documentclass[11pt]{article}
\usepackage[truedimen,margin=33mm]{geometry} 
\usepackage{mathrsfs}
\usepackage{amssymb}
\usepackage{color}
\usepackage{amsmath}
\usepackage{ascmac}
\usepackage{amsthm}
\usepackage[dvips]{graphicx}
\usepackage{booktabs}

\newtheorem{thm}{Theorem}
\newtheorem{lem}{Lemma}

\newtheorem{algo}{Algorithm}
\newtheorem{as}{Assumption}
\def\be{{\beta}}

\def\de{{\delta}}
\def\ep{{\varepsilon}}
\def\la{{\lambda}}

\def\th{{\theta}}

\def\bbe{{\text{\boldmath $\beta$}}}

\def\bOm{{\text{\boldmath $\Omega$}}}

\def\bphi{{\text{\boldmath $\phi$}}}

\def\beh{{\widehat \be}}

\def\thh{{\widehat \th}}

\def\muh{{\widehat \mu}}
\def\lah{{\widehat \la}}

\def\nuh{{\widehat \nu}}

\def\mh{\widehat{m}}
\def\phih{{\widehat \phi}}

\def\lat{{\widetilde \la}}
\def\mut{{\widetilde \mu}}

\def\bbeh{{\widehat \bbe}}

\def\bphih{{\widehat \bphi}}

\def\0{{\text{\boldmath $0$}}}

\def\e{{\text{\boldmath $e$}}}

\def\x{{\text{\boldmath $x$}}}

\def\B{{\text{\boldmath $B$}}}

\def\E{{\text{\boldmath $E$}}}

\def\Re{\mathbb{R}}
\def\Var{{\rm Var}}

\def\E{{\rm E}}
\def\ph{\widehat{p}}

\def\mh{\widehat{m}}
\def\Bh{\widehat{\B}}
\def\bOmh{\widehat{\bOm}}
\def\tr{{\rm tr}}

\def\Ga{\Gamma}

\title{{\bf Empirical Uncertain Bayes Methods\\ in Area-level Models\footnote{This version : \today}}}

\date{}

\begin{document}

\maketitle

\vspace{-0.9cm}\noindent
SHONOSUKE SUGASAWA\\
{\it Risk Analysis Research Center, The Institute of Statistical Mathematics}\\
TATSUYA KUBOKAWA\\
{\it Faculty of Economics, University of Tokyo}\\
KOTA OGASAWARA\\
{\it Department of Industrial Engineering and Economics, Tokyo Institute of Technology}\\

\vspace{0.8cm}\noindent
{\large\bf Abstract.}\ \
Random effects model can account for the lack of fitting a regression model and increase precision of estimating area-level means.
However, in case that the synthetic mean provides accurate estimates, the prior distribution may inflate an estimation error.
Thus it is desirable to consider the uncertain prior distribution, which is expressed as the mixture of a one-point distribution and a proper prior distribution.
In this paper, we develop an empirical Bayes approach for estimating area-level means, using the uncertain prior distribution in the context of a natural exponential family, which we call the empirical uncertain Bayes (EUB) method.
The regression model considered in this paper includes the Poisson-gamma and the binomial-beta, and the normal-normal (Fay-Herriot) model, which are typically used in small area estimation.
We obtain the estimators of hyperparameters based on the marginal likelihood by using a well-known Expectation-Maximization algorithm, and propose the EUB estimators of area means.
For risk evaluation of the EUB estimator, we derive a second-order unbiased estimator of a conditional mean squared error by using some techniques of numerical calculation.
Through simulation studies and real data applications, we evaluate a performance of the EUB estimator and compare it with the usual empirical Bayes estimator.

\vspace{0.5cm}\noindent

%\vspace{0.5cm}\noindent
%{\it Running headline}: Empirical uncertain Bayes method

%-------------------------------------------------------------------------------------------------------------------------------
%    SECTION   ---INTRODUCTION---
%-------------------------------------------------------------------------------------------------------------------------------
\section{Introduction}
In the small area estimation methodology, the model-based empirical Bayes approach has become very popular to produce indirect estimates since a design-based direct estimate of a small area characteristic based on the area sample is usually less reliable because the size of the area sample is small.
Importance and usefulness of model-based small area estimation approach may be assessed from an explosive growth of research publications. 
For a recent comprehensive review of this literature, see Rao and Molina (2015), Jiang and Lahiri (2006), and Pfeffermann (2013).
Random effects models can account for the lack of fit of a regression model and increase precision of estimation for small area mean.
The representative model for estimating area means is the Fay-Herriot model (Fay and Herriot, 1979) described as
$$
y_i=\theta_i+\ep_i, \ \ \ \th_i=\x_i^t\bbe+v_i, \ \ \ \ \ i=1,\ldots,m,
$$
where $\theta_i$ corresponds to an area mean, a parameter of interest, $\ep_i$ is a sampling error distributed as $\ep_i\sim N(0,D_i)$ for known $D_i$, and $v_i$ is a random effect distributed as $v_i\sim N(0,A)$ with unknown scalar parameter $A$.

Recently, however, Datta, Hall and Mandal (2011) showed that the model parameters and the small area means may be estimated with substantially higher accuracy if the random effects can be dispensed with via a suitable test.
In response to this work, Datta and Mandal (2015) proposed the Fay-Herriot model with uncertain random effects:
\begin{equation}\label{ureFH}
y_i=\theta_i+\ep_i, \ \ \ \th_i=\x_i^t\bbe+s_iv_i, \ \ \ \ \ i=1,\ldots,m,
\end{equation}
for $P(s_i=1)=p=1-P(s_i=0)$.
In Datta and Mandal (2015), the term $s_iv_i$ is called ``uncertain random effect" since the distribution of $s_iv_i$ is expressed as a mixture of $N(0,A)$ and the one-point distribution on $0$.
In this setting, the marginal density of $y_i$ is
$$
f(y_i)=\frac{1}{\sqrt{2\pi}}\left\{\frac{p}{\sqrt{A+D_i}}\exp\left(-\frac{(y_i-\x_i^t\bbe)^2}{2(A+D_i)}\right)+\frac{1-p}{\sqrt{D_i}}\exp\left(-\frac{(y_i-\x_i^t\bbe)^2}{2D_i}\right)\right\},
$$
which is a mixture density of the marginal density of the usual Fay-Herriot model and the simple regression model without random effects.
The resulting Bayes estimator (predictor) of $\theta_i$ is given by
$$
\thh_i=\x_i^t\bbe+\frac{r_iA}{A+D_i}(y_i-\x_i^t\bbe),\ \ \ \ r_i=p\left\{p+(1-p)\sqrt{\frac{A+D_i}{D_i}}\exp\left(-\frac{A(y_i-\x_i^t\bbe)^2}{2D_i(A+D_i)}\right)\right\}^{-1}.
$$
It is noted that if $p=1$, it follows $r_i=1$ in which $\thh_i$ corresponds to the usual linear estimator of $\th_i$.
It is worth noting that the shrinkage coefficient $r_iA/(A+D_i)$ depends on $y_i$ as well as $D_i$, and $r_i$ increases as the squared residual $(y_i-\x_i^t\bbe)^2$ increases.
Thus if the squared residual is small, which corresponds to the situation that $y_i$ is well explained by $\x_i$, the shrinkage coefficient gets smaller and $\thh_i$ becomes close to the synthetic mean $\x_i^t\bbe$.
Therefore, the use of the uncertain random effect enables us to take the information about how much $\x_i$ can explain $y_i$ into account for estimating $\th_i$.

While Datta and Mandal (2015) focused on extending the traditional Fay-Herriot model, we often encounter count data in real applications, and the model based on normal distributions is apparently inappropriate in such a case.
Thus, in this paper, we develop a method in a family of distributions called the natural exponential family with quadratic variance functions (NEF-QVF), which includes a normal distribution as well as binomial and Poisson distributions as a special case.
This family of distributions was originally investigated in Morris (1982, 1983), and the small area models based on this family were developed by Ghosh and Maiti (2004).
To implement the idea of uncertain random effects in this context, we rewrite the uncertain random effect model (\ref{ureFH}) as the hierarchical form:
\begin{align*}
&y_i|\th_i\sim N(0,D_i), \ \ \ \th_i|(s_i=1)\sim N(\x_i^t\bbe,A), \ \ \ \ \th_i|(s_i=0)=\x_i^t\bbe.
\end{align*}
This means that one of the two distributions $N(\x_i^t\bbe,A)$ and the one point distribution on $\x_i^t\bbe$ is randomly selected for the prior distribution of $\th_i$ in each area. 
In this paper, we naturally extend the idea to the NEF-QVF family.
Since there exist the conjugate priors for the natural parameter $\th_i$, we introduce the uncertain prior for $\th_i$, a mixture distribution of the conjugate prior and the one-point distribution on the synthetic mean.

For estimating area means under the model, we here develop an empirical Bayes (EB) approach while Datta and Mandal (2015) considered a hierarchical Bayes (HB) approach.
In the normal case as considered in Datta and Mandal (2015), a full Bayesian approach is relatively attractive since all the full conditional distribution of the model parameters as well as the random effects have familiar forms, so that we can efficiently sample from the posterior distribution using a Gibbs sampling.
However, in the non-normal case, the posterior distribution of the model parameters are not necessarily in familiar forms, so that we need to rely on an inefficient sampling algorithm such as a Metropolis-Hastings algorithm.
Moreover, the HB approach requires checking prior sensitivity and monitoring the convergence of the MCMC algorithm.
As suggested in Datta and Mandal (2015), the use of non-informative (improper) enables us to avoid subjective specification of priors, but the posterior propriety is not straightforward under non-normal cases. 
On the other hand, the empirical Bayes approach can enjoy easily computing point estimates of model parameters and Bayes estimator without requiring prior distributions. 
Since one of the greatest purposes in small area estimation is point estimation, the EB approach is more attractive in this case.

Owing to the conjugacy of the prior distribution, we can easily establish the Expectation-Maximization (EM) algorithm for maximizing the marginal likelihood function to get the estimates of model parameters.
Using the estimator, we derive the the empirical uncertain Bayes (EUB) estimator of the area mean.
For calibration of uncertainty of the EUB estimator, we consider the conditional mean squared error (CMSE) and derive the second-order unbiased estimator motivated from the work of Booth and Hobert (1998), Datta, Kubokawa, Molina and Rao (2011) and Sugasawa and Kubokawa (2016).
As typical examples, we handle three models, namely the Fay-Herriot model for continuous data, the  Poisson-gamma and binomial-beta models for count data.
It is shown that the shrinkage property pointed out in Datta and Mandal (2015) in the Fay-Herriot model still holds in both the Poisson-gamma and binomial-beta models.
That is, the shrinkage coefficient in the EUB estimator decreases as the $y_i$ gets close to the synthetic mean.

This paper is organized as follows. 
In Section \ref{sec:model}, we provide the detailed description of the proposed model, the EM algorithm for parameter estimation and three examples.
In Section \ref{sec:mse}, we derive the second order unbiased estimator of CMSE for risk evaluation of the EUB estimator.
Simulation studies and empirical applications are given in \ref{sec:sim} and \ref{sec:application}, respectively.

%-------------------------------------------------------------------------------------------------------------------------------
%    ---SECTION---     MODEL
%-------------------------------------------------------------------------------------------------------------------------------
\section{Empirical Uncertain Bayes Methods}\label{sec:model}

%-------------------------------------------------------------------------------------------------------------------------------
%    SETTINGS
%-------------------------------------------------------------------------------------------------------------------------------
\subsection{Model setup and uncertain Bayes estimator}
Let $y_1, \ldots, y_m$ be mutually independent random variables where the conditional distribution of $y_i$ given $\th_i$ belongs to the the following natural exponential family:
\begin{equation}\label{obs}
y_i | \th_i \sim f(y_i|\th_i)=\exp\{n_i(\th_i y_i -\psi(\th_i)) + c(y_i,n_i)\},
\end{equation}
where $n_i$ is a known scalar parameter and is not necessarily corresponding to the sample size in the $i$th area.
As the prior distribution of $\theta_i$, we set the uncertain random structure treated in Datta and Mandal (2015).
Let $s_1, \ldots, s_m$ be mutually independent and identical random variables distributed as 
$$
P(s_i=1)=p, \ \ \ P(s_i=0)=1-p.
$$
The prior distribution of $\theta_i$ is given by
\begin{equation}\label{prior}
\begin{split}
\th_i|(s_i=1)\sim \pi(\th_i) = \exp\left\{\nu(m_i\th_i-\psi(\th_i))+C(\nu,m_i)\right\}, \ \ \ \ \th_i|(s_i=0)=(\psi')^{-1}(m_i),
\end{split}
\end{equation}
where $\nu$ is an unknown scalar hyperparameter, $C(\nu,m_i)$ is the normalizing constant and  $\psi'(t)=d\psi(t)/dt$.
In our settings, we consider the canonical link
$$
m_i=\psi'(\x_i^t\bbe), 
$$
where $\x_i$ is a $q\times 1$ vector of explanatory variables, $\bbe$ is a $q\times 1$ unknown common vector of regression coefficients.
The function $f(y_i|\th_i)$ is the regular one-parameter exponential family and the function $\pi(\th_i)$ is the conjugate prior distribution.
Then the unknown parameter in two-stage model (\ref{obs}) and (\ref{prior}) are $\bphi=(\bbe^t,\nu,p)^t$.
The quantity of interest in this paper is the conditional expectation of $y_i$ given $\th_i$, defined as 
$$
\mu_i =\E[y_i|\th_i] = \psi'(\th_i),
$$
noting that $\mu_i|(s_i=0)=m_i$ from (\ref{prior}).
For $\psi''(t)=d^2\psi(t)/dt^2$, we assume that $\psi''(\th_i)=Q(\mu_i)$, namely, 
$$
\Var(y_i|\th_i)={\psi''(\th_i)\over n_i}={Q(\mu_i)\over n_i},
$$
with $Q(x)=v_0+v_1 x+v_2x^2$ for known constants $v_0$, $v_1$ and $v_2$ which are not simultaneously zero.
This means that the conditional variance $\Var(y_i|\theta_i)$ is a quadratic function of the conditional expectation $\E[y_i|\th_i]$.
This family is known as a natural exponential family with the quadratic variance function studied by Morris (1982, 1983).
Similarly, the mean and variance of the prior distribution given $s_i=1$ are
\begin{equation*}
\E[\mu_i|s_i=1] =m_i, \quad \Var(\mu_i|s_i=1)={Q(m_i)\over \nu - v_2}.
\end{equation*}
The joint density (or mass) function of $(y_i,\theta_i,s_i)$ is 
\begin{align*}
g(y_i,\th_i,s_i=1)=f(y_i|\th_i)\pi(\th_i), \ \ \ \ g(y_i,\th_i,s_i=0)=\delta_{\th_i}((\psi')^{-1}(m_i))f(y_i|\th_i),
\end{align*}
where $\de_{\th_i}(a)$ denotes the point mass at $\th_i=a$.
Then the joint distribution of $(y_i,\th_i)$ and the marginal distribution of $y_i$ are both mixtures of two distributions:
\begin{align*}
g(y_i,\th_i)&=pf(y_i|\th_i)\pi(\th_i)+(1-p)\delta_{\th_i}((\psi')^{-1}(m_i))f(y_i|\th_i),
\\
f(y_i;\bphi)&=pf_1(y_i; \bphi)+(1-p)f_2(y_i; \bphi),
\end{align*}
where
\begin{equation}\label{marg}
f_1(y_i; \bphi)=\int f(y_i|\th_i)\pi(\th_i)d\th_i, \ \ \ \ f_2(y_i;\bphi)=f(y_i|\th_i=(\psi')^{-1}(m_i)).
\end{equation}
Since $\pi(\th_i)$ is the conjugate prior of $\th_i$, the marginal distribution $f_1(y_i; \bphi)$ and the conditional distribution $\pi(\th_i|y_i,s_i=1; \bphi)$ can be obtain in the closed forms:
\begin{align*}
\pi(\th_i|y_i, s_i=1;\bphi)&=\exp\big\{(n_i+\nu)(\eta_i\th_i-\psi(\th_i))\big\}C(n_i+\nu, \eta_i), \\
f_1(y_i; \bphi)&=\frac{C(\nu,m_i)}{C(n_i+\nu,\eta_i)}\exp\big\{c(y_i, n_i)\big\},
\end{align*}
where 
$$
\eta_i\equiv \eta_i(y_i; \bphi)=\frac{n_iy_i+\nu m_i}{n_i+\nu}.
$$
The conditional distribution of $s_i$ given $y_i$ can be obtained as
$$
P(s_i=1|y_i;\bphi)=\frac{pf_1(y_i,\bphi)}{f(y_i;\bphi)}=\frac{p}{p+(1-p)f_2(y_i; \bphi)/f_1(y_i;\bphi)}=1-P(s_i=0|y_i;\bphi).
$$
To obtain the Bayes estimator of $\mu_i$, we note that 
\begin{equation}\label{cond.B}
\E\left[\mu_i|s_i,y_i;\bphi\right]=m_i+\frac{n_i}{\nu+n_i}(y_i-m_i)I(s_i=1),
\end{equation}
where $I(\cdot)$ is an indicator function.
Hence the Bayes estimator of $\mu_i$ is 
\begin{equation}\label{UB}
\begin{split}
\mut_i(y_i,\bphi)=\E\left[\mu_i|y_i;\bphi\right]=\E\left[\E\left(\mu_i|s_i,y_i;\bphi\right)|y_i;\bphi\right]&=m_i+\frac{n_i}{\nu+n_i}(y_i-m_i)r_i(y_i,\bphi),
\end{split}\end{equation}
where 
\begin{equation}\label{py}
r_i(y_i,\bphi)=P(s_i=1|y_i;\bphi)=\frac{p}{p+(1-p)f_2(y_i; \bphi) / f_1(y_i; \bphi)}.
\end{equation}
It is observed that $r_i(y_i,\bphi)$ increases in $p$ and decreases in the ratio $f_2(y_i;\bphi)/f_1(y_i;\bphi)$.
In what follows, we use the abbreviated notations $\mut_i$ and $r_i$ instead of $\mut_i(y_i,\bphi)$ and $r_i(y_i,\bphi)$, respectively, when there is no confusion.
It is noted that the Bayes estimator (\ref{UB}) can be expressed as 
$$
\mut_i=y_i-\left\{1-\frac{n_i}{\nu+n_i}r_i(y_i,\bphi)\right\}(y_i-m_i),
$$
which shrinks the direct estimator $y_i$ toward the regression (or synthetic) part $m_i=\psi'(\x_i^t\bbe)$, and the shrinkage function depends on $y_i$ through $r_i$.
On the other hand, in the classical two-stage model as used in Ghosh and Maiti (2004), the shrinkage function does not depends on the observation $y_i$, which is sometimes not flexible for real data analysis.
It should be noted that $r_i=1$ when $p=1$.
Thus, the suggested method includes the classical method as well as it has the shrinkage function adjusted by $y_i$ which arises from introducing the weight parameter $p$.
Moreover, when the prior is completely singular, namely $p=0$, it follows $r_i=0$ and the resulting Bayes estimator is $m_i$.
We call the estimator (\ref{UB}) under the prior (\ref{prior}) the uncertain Bayes estimator in order to distinguish from the conventional Bayes estimator.

%-------------------------------------------------------------------------------------------------------------------------------
%    Parameter estimation
%-------------------------------------------------------------------------------------------------------------------------------
\subsection{Maximum likelihood estimation using EM algorithm}\label{sec:EM}
Since the uncertain Bayes estimator (\ref{UB}) depends on the unknown model parameter $\bphi$, we need to estimate them for practical use.
A reasonable method is the maximum likelihood (ML) estimator which maximizes the marginal distribution of $y_i$.
Since the mariginal density is the mixture of the two distributions $f(y_i; \bphi)=pf_1(y_i; \bphi)+(1-p)f_2(y_i; \bphi)$, the ML estimator is the maximizer of the log-likelihood function
\begin{equation}\label{est}
L(\bphi)=\sum_{i=1}^m\log\left\{pf_1(y_i; \bphi)+(1-p)f_2(y_i; \bphi)\right\}.
\end{equation} 
To compute the ML estimate, we propose Expectation-Maximization (EM) algorithm (Dempster, Laird and Rubin, 1977) which maximizes the objective function (\ref{est}) iteratively and indirectly.
From (\ref{obs}) and (\ref{prior}), the complete log-likelihood function $L^c(\bphi)$ given $(y_i,\theta_i,s_i)_{i=1,\ldots,m}$ is 
$$
L^c(\bphi)=\sum_{i=1}^m \left\{n_i(\th_iy_i-\psi(\th_i))\right\}
+\sum_{i=1}^ms_i\left\{\nu(m_i\theta_i-\psi(\theta_i))+C(\nu,m_i)\right\}
+\sum_{i=1}^m\{s_i\log p+(1-s_i)\log(1-p)\}.
$$
In the $r$th iteration, we first compute the expectation of the complete log-likelihood $\E^{(r)}[L^c(\bphi)]$ at the E-step, where $E^{(r)}$ denotes the expectation with respect to the conditional distributions  $(\th_i,s_i)|y_i$ with hyperparameter values $\bphi^{(r)}$.
Then the objective function to be maximized at the M-step in the $r$th iteration is 
\begin{equation*}
\begin{split}
Q^{(r)}(\bphi)\equiv\E^{(r)}[L^c(\bphi)]
&=\sum_{i=1}^mr_i(y_i,\bphi^{(r)})\left\{\nu m_i\E^{(r)}[\theta_i|s_i=1]-\nu\E^{(r)}[\psi(\theta_i)|s_i=1]+C(\nu,m_i)\right\}\\
& \ \ \ \ \ +\sum_{i=1}^m\Big\{r_i(y_i,\bphi^{(r)})\log p+(1-r_i(y_i,\bphi^{(r)}))\log(1-p)\Big\},
\end{split}
\end{equation*}
which yields the updating algorithm as
\begin{equation}\label{M-step}
\begin{split}
(\bbe^{(r+1)},\nu^{(r+1)})
&={\rm argmax}_{\bbe,\nu}\sum_{i=1}^mr_i(y_i,\bphi^{(r)})\left\{\nu m_i\E^{(r)}[\theta_i|s_i=1]-\nu\E^{(r)}[\psi(\theta_i)|s_i=1]+C(\nu,m_i)\right\}\\
p^{(r+1)}
&=\frac1m\sum_{i=1}^mr_i(y_i,\bphi^{(r)}).
\end{split}
\end{equation} 
Since the prior distribution of $\th_i$ given $s_i=1$ is conjugate, the posterior distribution of $\th_i$ given $s_i=1$ belongs to the same family as the prior distribution, and we can easily generate samples from the distribution in common models as demonstrated in the subsequent section.
Hence, the calculation of two expectations $\E^{(r)}[\theta_i|s_i=1]$ and $\E^{(r)}[\psi(\theta_i)|s_i=1]$ given in the E-step is easy to carry out.
We summarize the EM algorithm in the following.

\begin{algo}[EM algorithm] \ \ Iterative,
\begin{itemize}
\item[1.]
Set the initial value $\bphi^{(0)}$ and $r=0$. 

\item[2.]
Compute $\E^{(r)}[\theta_i|s_i=1]$ and $\E^{(r)}[\psi(\theta_i)|s_i=1]$ using the current parameter value $\bphi^{(r)}$.

\item[3.]
Update the parameter value as  $\bphi^{(r+1)}$ based on {\rm (\ref{M-step})}.

\item[4.]
If the difference between $\bphi^{(r)}$ and $\bphi^{(r+1)}$ is sufficiently small, then the estimate is given by $\bphi^{(r+1)}$. Otherwise, set $r=r+1$ and go back to Step 2. 
\end{itemize} 
\end{algo}

Finally, substituting $\bphih$ into the UB estimator, we get the empirical uncertain Bayes (EUB) estimator 
\begin{equation}\label{EUB}
\muh_i\equiv \mut_i(y_i,\bphih)
=\mh_i+\frac{n_i}{\nuh+n_i}(y_i-\widehat{m}_i)r_i(y_i, \bphih),
\end{equation}
where $\mh_i=\psi'(\x_i^t\bbeh)$.

%-------------------------------------------------------------------------------------------------------------------------------
%    EXAMPLES
%-------------------------------------------------------------------------------------------------------------------------------
\subsection{Some examples}\label{sec:example}
Here we provide three typical models often used in practice and investigate properties of the UB estimators with detailed expressions of E-step and M-step in the EM algorithm.

\ \\
{\bf [1] Normal-normal (Fay-Herriot) model}.\ \ \ 
 The Fay-Herriot model (Fay and Herriot, 1979) is an area-level model frequently used in small area estimation, given by
$$
y_i|\th_i\sim N(\th_i, D_i), \ \ \ \ \th_i|(s_i=1)\sim N(\x_i^t\bbe,A), \ \ \ \ i=1,\ldots,m,
$$
corresponding to $n_i=D_i^{-1}, v_0=1, v_1=v_2=0, \nu=A^{-1}$ and $\psi(\theta_i)=\theta_i^2/2$ in (\ref{obs}) and (\ref{prior}).
This model was studied in Datta and Mandal (2015) in terms of Bayesian perspectives.
The marginal distributions of $f_1(y_i)$ and $f_2(y_i)$ in (\ref{marg}) are given by
\begin{equation}\label{FH:dens}
f_1(y_i;\bphi)=\frac{1}{\sqrt{2\pi (A+D_i)}}\exp\left(-\frac{(y_i-m_i)^2}{2(A+D_i)}\right), \ \ \ \ f_2(y_i;\bphi)=\frac{1}{\sqrt{2\pi D_i}}\exp\left(-\frac{(y_i-m_i)^2}{2D_i}\right),
\end{equation}
so that $r_i(y_i,p)$ is obtained from (\ref{py}) as
$$
r_i(y_i,p)=p\left\{p+(1-p)\sqrt{\frac{A+D_i}{D_i}}\exp\left(-\frac{A(y_i-m_i)^2}{2D_i(A+D_i)}\right)\right\}^{-1},
$$
which coincides with the result given in Datta and Mandal (2015).
It is clear that $r_i(y_i,p)$ takes small values when $y_i$ is close to $m_i$, corresponding to the case where $y_i$ is well explained by $m_i$ without random effects.

Regarding the parameter estimation via the EM algorithm in the Fay-Heriot model, the objective function at the M-step is 
\begin{align*}
Q^{(r)}(\bbe,A)=-\frac12\sum_{i=1}^mr_i^{(r)}\left\{\log A+\frac1A\left(\th_i^{(r)}-\x_i^t\bbe\right)^2\right\},
\end{align*}
where $r_i^{(r)}=r_i(y_i,\bbe^{(r)},A^{(r)},p^{(r)})$ and $\th_i^{(r)}=(A^{(r)}y_i+D_i\x_i^t\bbe^{(r)})/(A^{(r)}+D_i)$, so that the updating step for $\bbe$ and $A$ is written as
$$
\bbe^{(r+1)}=\left(\sum_{i=1}^mr_i^{(r)}\x_i\x_i^t\right)^{-1}\sum_{i=1}^mr_i^{(r)}\th_i^{(r)}\x_i, \ \ \ \ 
A^{(r+1)}=\frac1m\sum_{i=1}^m\left(\th_i^{(r)}-\x_i^t\bbe^{(r+1)}\right)^2.
$$

\ \\
{\bf [2] Poisson-gamma model}. \ \ \ 
Let $z_1,\ldots,z_m$ be mutually independent random variables having
$$
z_i|\la_i\sim{\rm Po}(n_i\la_i), \ \ \ \ \ \la_i|(s_i=1)\sim{\rm Ga}(\nu m_i,\nu)
$$
where $\la_1,\ldots,\la_m$ are mutually independent, Po$(\la)$ denotes the Poisson distribution with mean $\la$, and Ga$(a,b)$ denotes the gamma distribution with density $f(x)\propto x^{a-1}\exp(-bx)$.
Let $y_i=z_i/n_i$ and $\log m_i=\x_i^t\bbe$ for $i=1,\ldots,m$. 
Then, the notations in (\ref{obs}) and (\ref{prior}) correspond to $v_1=1, \ v_0=v_2=0$ and $\psi(\theta_i)=\exp(\theta_i)$. 
The marginal distributions of $f_1(y_i)$ and $f_2(y_i)$ are given by
\begin{equation}\label{PG:dens}
f_1(y_i;\bphi)=\frac{\Gamma(n_iy_i+\nu m_i)}{\Gamma(n_iy_i+1)\Gamma(\nu m_i)}\left(\frac{n_i}{n_i+\nu}\right)^{n_iy_i}\left(\frac{\nu}{n_i+\nu}\right)^{\nu m_i},\ \ \ \ 
f_2(y_i;\bphi)=\frac{(n_im_i)^{n_iy_i}}{(n_iy_i)!}\exp(-n_im_i)
\end{equation}
where $\Gamma(\cdot)$ denotes a gamma function, so that $r_i(y_i,p)$ is written as
\begin{equation}\label{PG-prob}
r_i(y_i,p)=p\left\{p+(1-p)\frac{\Ga(\nu m_i)\exp(-n_im_i)}{\Ga(n_iy_i+\nu m_i)}(n_i+\nu)^{n_iy_i+\nu m_i}m_i^{n_iy_i}\nu^{-\nu m_i}\right\}^{-1}.
\end{equation}
Unlike the Fay-Herriot model, it is not clear when $r_i(y_i,p)$ takes small values as a function of $y_i$.
To see this property, let $h(z_i)=(n_i+\nu)^{z_i+\nu m_i}m_i^{z_i}/\Gamma(z_i+\nu m_i)$.
It is noted that $r_i(y_i,p)$ depends on $z_i(=n_iy_i)$ through $h(z_i)$.
It follows that 
$$
\frac{h(z_i+1)}{h(z_i)}=\frac{n_im_i+\nu m_i}{z_i+\nu m_i},
$$
so that we have $h(z_i)\leq h(z_i+1)$ for $y_i\leq m_i$ and $h(z_i)\geq h(z_i+1)$ for $y_i\geq m_i$.
Then, when $y_i$ is close to $m_i$, $h(z_i)$ takes a large value, which results in a small value of   $r_i(y_i,p)$.
This observation is　similar to the case of the Fay-Herriot model.

The objective function at the M-step in the EM algorithm can be expressed as 
\begin{align*}
Q^{(r)}(\bbe,\nu)=\sum_{i=1}^m&r_i(y_i,\bbe^{(r)},\nu^{(r)},p^{(r)})\bigg\{\nu m_i\log\nu-\log\Ga(\nu m_i)\\
&+\nu m_i\int_0^{\infty}(\log t) f_{\Ga}(t; n_iy_i+\nu^{(r)} m_i^{(r)},n_i+\nu^{(r)})dt-\nu\frac{n_iy_i+\nu^{(r)}m_i^{(r)}}{n_i+\nu^{(r)}}\bigg\},
\end{align*}
where $f_{\Ga}(\cdot; a,b)$ denotes the density function of ${\rm Ga}(a,b)$. 
 It should be noted that the integral given in the objective function can be easily calculated by generating samples from ${\rm Ga}(n_iy_i+\nu^{(r)} m_i^{(r)},n_i+\nu^{(r)})$.

\ \\
{\bf [3] Binomial-beta model.} \ \ \ 
Let $z_1,\ldots,z_m$ be mutually independent random variables having
$$
z_i|p_i\sim {\rm Bin}(n_i,p_i), \ \ \ p_i|(s_i=1)\sim {\rm Beta}(\nu m_i,\nu(1-m_i)),
$$
where $p_1,\ldots,p_m$ are mutually independent, Bin$(n,p)$ denotes the binomial distribution and Beta$(a,b)$ denotes the beta distribution with density $f(x)\propto x^{a-1}(1-x)^{b-1}$. 
Let $y_i=z_i/n_i$ and $m_i=\exp(\x_i^t\bbe)/(1+\exp(\x_i^t\bbe))$ for $i=1,\ldots,m$. 
Then the notations in (\ref{obs}) and (\ref{prior}) correspond to $v_0=0, \ v_1=1$ and $v_2=-1, \ \mu_i=p_i=\exp(\theta_i)/(1+\exp(\theta_i))$ and $\psi(\theta_i)=\log(1+\exp(\theta_i))$. 
The marginal distributions of $f_1(y_i)$ and $f_2(y_i)$ are 
\begin{equation*}\label{BB:dens}
f_1(y_i;\bphi)=\binom{n_i}{n_iy_i}\frac{B(\nu m_i+n_i y_i, n_i(1-y_i)+\nu(1-m_i))}{B(\nu m_i,\nu(1-m_i))},\ \ \ \ \ 
f_2(y_i;\bphi)=\binom{n_i}{n_iy_i}m_i^{n_iy_i}(1-m_i)^{n_i(1-y_i)},
\end{equation*}
where $B(\cdot, \cdot)$ denotes a beta function, so that $r_i(y_i,p)$ is written as
$$
r_i(y_i,p)=p\left\{p+(1-p)\frac{B(\nu m_i,\nu(1-m_i))}{B(\nu m_i+n_i y_i, n_i(1-y_i)+\nu(1-m_i))}m_i^{n_iy_i}(1-m_i)^{n_i(1-y_i)}\right\}^{-1}.
$$
Using the same arguments as in the Poisson-gamma model, we consider the function $h(z_i)=m_i^{z_i}(1-m_i)^{n_i-z_i}/B(\nu m_i+z_i,n_i-z_i+\nu(1-m_i))$.
Then the straightforward calculation shows that
$$
\frac{h(z_i+1)}{h(z_i)}=\frac{m_i\left\{n_i-z_i-1+\nu(1-m_i)\right\}}{(1-m_i)(\nu m_i+z_i)},
$$
whereby $h(z_i)\leq h(z_i+1)$ for $y_i\leq m_i(1-n_i^{-1})$ and $h(z_i)\geq h(z_i+1)$ for $y_i\geq m_i(1-n_i^{-1})$.
Thus, when $y_i$ is close to $m_i$, $h(z_i)$ takes a large value, which results in a small value of   $r_i(y_i,p)$.

The objective function at the M-step in the EM algorithm is expressed as 
\begin{align*}
Q^{(r)}(\bbe,\nu)=\sum_{i=1}^m&r_i(y_i,\bbe^{(r)},\nu^{(r)},p^{(r)})\bigg\{\nu m_i\int_0^1(\log t) f_{B}(t; a_i^{(r)},b_i^{(r)})dt\\
&+\nu(1-m_i)\int_0^1\log (1-t) f_{B}(t; a_i^{(r)},b_i^{(r)})dt-\log B(\nu m_i,\nu(1-m_i))\bigg\},
\end{align*}
where $a_i^{(r)}=n_iy_i+\nu^{(r)} m_i^{(r)}$, $b_i^{(r)}=n_i(1-y_i)+\nu^{(r)}(1-m_i^{(r)})$ and $f_{B}(\cdot; a,b)$ denotes the density function of the beta distribution ${\rm Beta}(a,b)$. 
The two integrals given in the above formula can be easily computed by generating samples from ${\rm Beta}(a_i^{(r)},b_i^{(r)})$.

\begin{figure}[htbp!]
\begin{center}
\includegraphics[width=4.9cm]{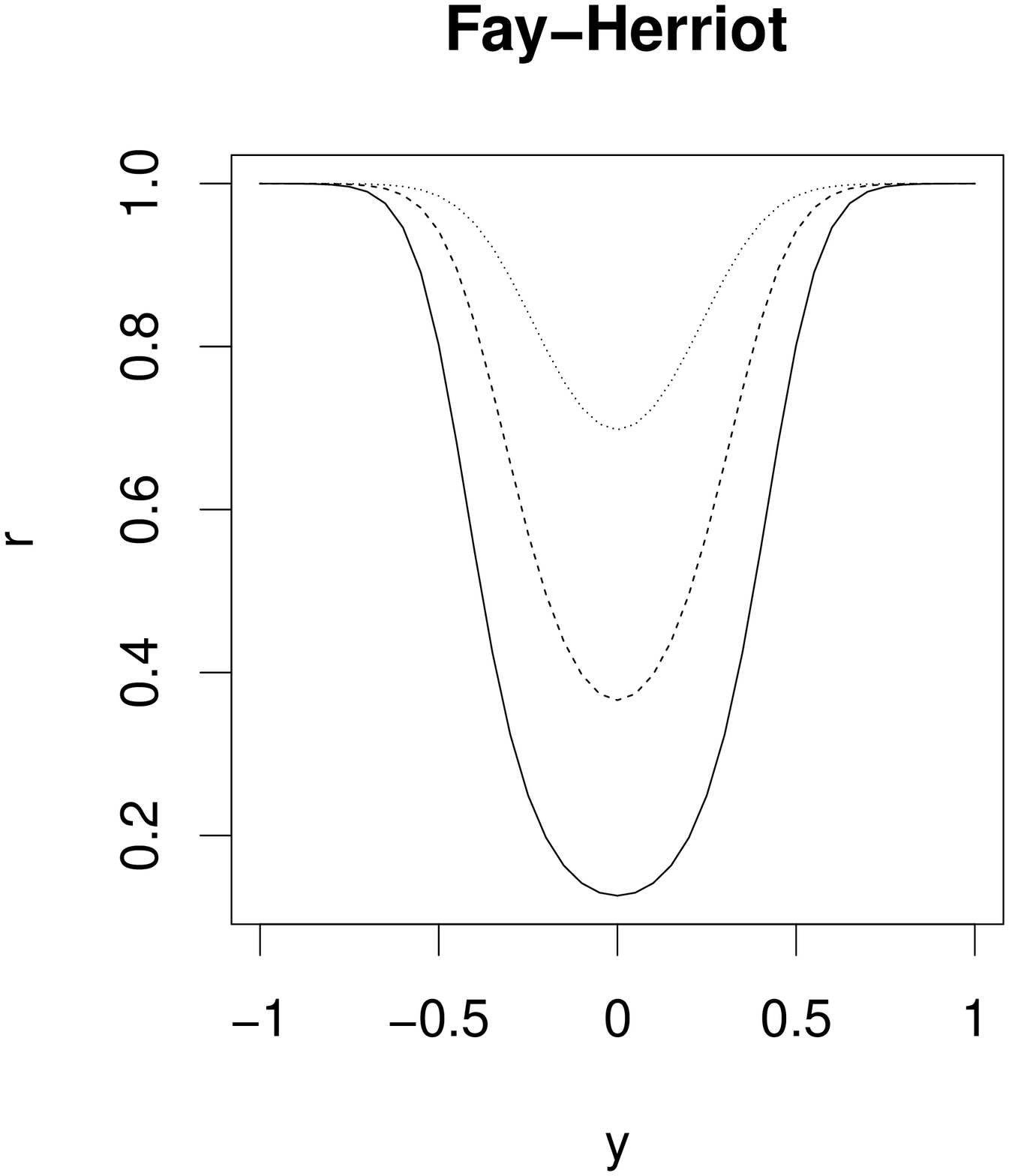}
\includegraphics[width=4.9cm]{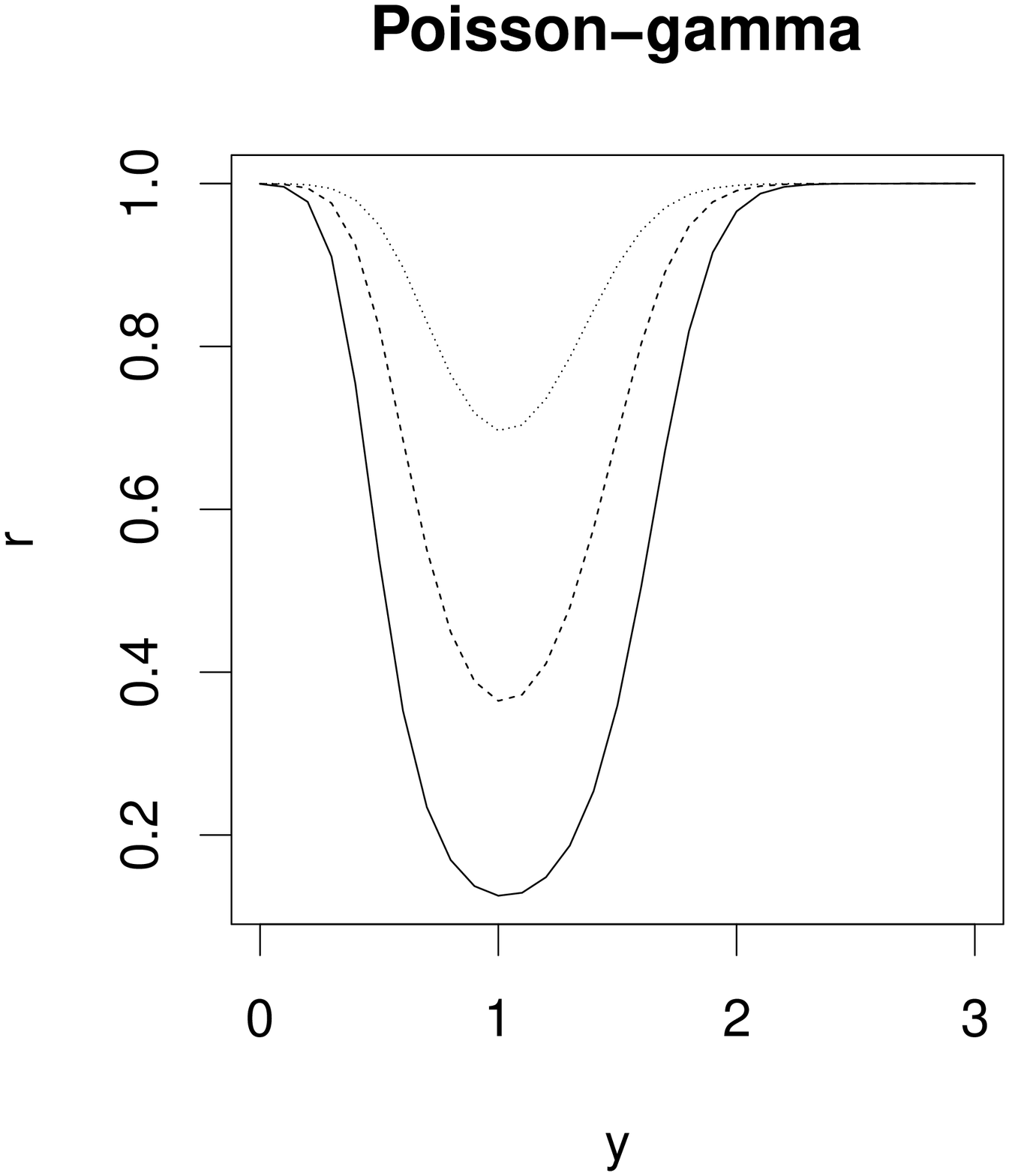} 
\includegraphics[width=4.9cm]{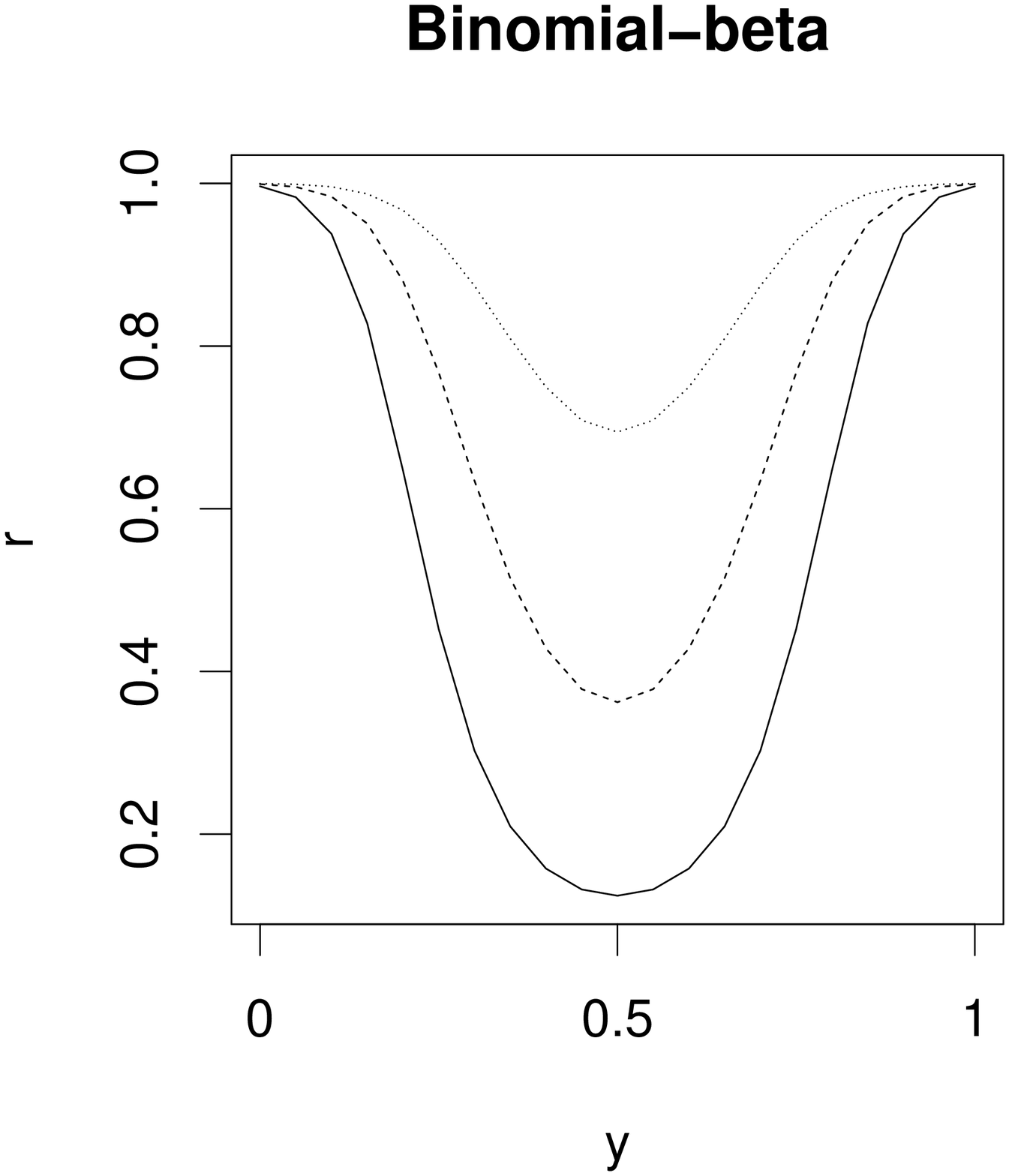}
\end{center}
\caption{Shrinkage function $r_i(y_i,p)$ in the thee models with $p=0.2$ (solid), $0.5$ (dashed), and $0.8$ (dotted).
\label{fig:r}
} 
\end{figure}

\vspace{0.7cm}
In Figure \ref{fig:r}, we draw the shrinkage function $r_i(y_i,p)$ as a function of $y_i$  for the three models, where $n_i=10$, $\nu=10$, $m_i=\psi'(\beta)$ at $\beta=0$, and the solid, dashed and dotted lines correspond to the three values $p=0.2$, $0.5$ and $0.8$, respectively.
It is observed from Figure \ref{fig:r} that the shrinkage function in all the models are actually minimized at $y_i=m_i$ as discussed so far, and converges to $1$ as $y_i$ goes away from $m_i$.
Especially, it is interesting to point out that in the Poisson-gamma model, the shrinkage ratio is not symmetric around $y_i=m_i$, while the other two models are symmetric around $y_i=m_i$.

% FIG  ---COMPARISON---

%-------------------------------------------------------------------------------------------------------------------------------
%    ---SECTION---  MSE
%-------------------------------------------------------------------------------------------------------------------------------
\section{Risk Evaluation of the EUB Estimator}\label{sec:mse}

%-------------------------------------------------------------------------------------------------------------------------------
%    MSE FORMULA
%-------------------------------------------------------------------------------------------------------------------------------
\subsection{Conditional MSE of the EUB estimator}
In practice, the risk evaluation of the resulting estimator is an important issue in small area estimation.
The unconditional mean squared error (MSE) is often used, but it is not suitable in this context, because researchers are interested in the risk of the area-specific risk in predicting $\mu_i$ under given $y_i$.
This philosophy was originally proposed by Booth and Hobert (1998), and they suggested using the conditional MSE (CMSE) instead of the classical unconditional MSE in the context of mixed model prediction.
Since then, the CMSE has been studied in the literature of small area estimation, including Datta, et al. (2011) and Sugasawa and Kubokawa (2016).
The CMSE of the EUB estimator is defined as 
$$
{\rm CM}_i(y_i,\bphi)=\E\left[(\muh_i-\mu_i)^2|y_i;\bphi\right],
$$
noting that the expectation is taken with respect to the conditional distribution $Y_{(-i)}|y_i$ with $Y_{(-i)}=\{y_1,\ldots,y_{i-1},y_{i+1},\ldots,y_m\}$.
Because $\mut_i$ is the conditional expectation, the CMSE can be decomposed as 
\begin{equation}\label{CMSE}
{\rm CM}_i=\Var(\mu_i|y_i;\bphi)+\E\left[(\muh_i-\mut_i)^2|y_i;\bphi\right].
\end{equation}
We shall evaluate the two terms in the right hand side of (\ref{CMSE}).

Concerning the first term in (\ref{CMSE}), owing to the quadratic variance structure of the assumed model, we have 
$$
\Var(\mu_i|y_i;\bphi)=\E\left[\Var(\mu_i|s_i,y_i;\bphi)|y_i;\bphi\right]+\Var\left(\E[\mu_i|s_i,y_i;\bphi]|y_i;\bphi\right).
$$
In the case of $s_i=1$, we have $\Var(\mu_i|s_i=1, y_i;\bphi)=Q(\eta_i)/(n_i+\nu-v_2)$ for $\eta_i=\E[\mu_i|s_i=1, y_i;\bphi]=(n_iy_i+\nu m_i)/(n_i+\nu)$.
Thus, 
\begin{align*}
\Var(\mu_i|s_i,y_i;\bphi)={Q(\eta_i) \over n_i+\nu-v_2} I(s_i=1).
\end{align*}
From (\ref{cond.B}), it follows that
\begin{equation}\label{R10}
\Var(\mu_i|y_i;\bphi)=
{Q(\eta_i) \over n_i+\nu-v_2}P(s_i=1|y_i;\bphi) + \Var\Big(m_i+{n_i\over\nu+n_i}(y_i-m_i)I(s_i=1)|y_i;\bphi\Big).
\end{equation}
Here it is observed that
\begin{align*}
\Var\Big(m_i+{n_i\over\nu+n_i}(y_i-m_i)I(s_i=1)|y_i;\bphi\Big)
=&\Big({n_i\over \nu+n_i}\Big)^2(y_i-m_i)^2 E\Big[ I(s_i=1)- 2r_i I(s_i=1)+ r_i^2 |y_i;\bphi\Big]
\\
=&\Big({n_i\over \nu+n_i}\Big)^2(y_i-m_i)^2 r_i(1-r_i),
\end{align*}
where $r_i$ is given in (\ref{py}).
We thus get
\begin{equation}\label{R1}
R_{1i}(y_i,\bphi)\equiv \Var(\mu_i|y_i;\bphi)=
\frac{n_i^2}{(\nu+n_i)^2}(y_i-m_i)^2r_i(1-r_i)+{r_iQ(\eta_i) \over n_i+\nu-v_2},
\end{equation}
which is of order $O_p(1)$.

Concerning the second term $\E\left[(\muh_i-\mut_i)^2|y_i\right]$ in (\ref{CMSE}), we approximate it up to second order.
For notational simplicity, let $\bphi=(\phi_1, \ldots,\phi_q, \phi_{q+1}, \phi_{q+2})^t$ for $(\phi_1, \ldots,\phi_q)^t=\bbe$, $\phi_{q+1}=\nu$ and $\phi_{q+2}=p$.
Let $\bphih=(\phih_1, \ldots, \phih_q, \phih_{q+1}, \phih_{q+2})^t$ be the ML estimator of $\bphi$, where $(\phih_1, \ldots,\phih_q)^t=\bbeh$, $\phih_{q+1}=\nuh$ and $\phih_{q+2}=\ph$. 
The asymptotic variance and bias of $\bphih$ are, respectively, written as 
$$
\bOm\equiv\E\left\{(\bphih-\bphi)(\bphih-\bphi)^t\right\}, \ \ \ \B\equiv\E\left(\bphih-\bphi\right).
$$
It is noted that $\bOm$ and $\B$ are of order $O(m^{-1})$.
Assume the following regularity conditions.

% Assumptons
\medskip
\begin{as}\label{as}
\item[{\rm (i)}]
There exist $\underline{n},\overline{n}>0$ such that $\underline{n}\leq n_i\leq\overline{n}$ for all $i=1,\ldots,m$.

\item[{\rm (ii)}]
The true value of the parameter $\bphi$ is in the interior of $\Phi$, where $\Phi$ is the parameter space.

\item[{\rm (iii)}]
The densities $f_{a}(y_i;\bphi)$ for $a=1,2$ are three times continuously differentiable and satisfies for $j,\ell,k=1,\ldots,q+2$,
$$
|f_{a(\phi_j)}(y_i;\bphi)|+|f_{a(\phi_j\phi_{\ell})}(y_i;\bphi)|+|f_{a(\phi_j\phi_{\ell}\phi_k)}(y_i;\bphi)|\leq C(y_i,\bphi),
$$
for fixed $\bphi$ and $\E[|C(y_i,\bphi)|^{4+\delta}]<\infty$ for some $\delta>0$.
\end{as}

\medskip
The assumption (i) is a standard one in this context. 
For example, in the Fay-Heriot model described in Section \ref{sec:example}, the assumption corresponds to $\underline{D}\leq D_i\leq\overline{D}, \ i=1,\ldots,m$ for some $\underline{D}$ and $\overline{D}$, which is usually assumed in the context of small area estimation (e.g. Datta, Rao and Smith, 2005).
The assumptions (ii) and (iii) are required for deriving the asymptotic properties of the ML estimator of $\bphi$ as provided in Lemma \ref{lem:asymp}.
It should be noted that the typical three models described in Section \ref{sec:example} satisfy the assumption (iii), which can be demonstrated in the Appendix.

Since $y_1,\ldots,y_m$ are mutually independent, from Theorem 1 in Lohr and Rao (2009), we can get the following lemma about asymptotic properties of estimators.

% LEMMA  ---parameter estimation---
\begin{lem}\label{lem:asymp}
For the ML estimator $\bphih$, under Assumption 1,  it holds that $\sqrt{m}(\bphih-\bphi)=O_p(1)$, $\E\left\{(\bphih-\bphi)(\bphih-\bphi)^t |y_i\right\}=\bOm+o_p(m^{-1})$ and
$$
\E(\bphih-\bphi|y_i)=\B-\bOm L_{i(\bphi)}(y_i,\bphi)+o_p(m^{-1}),
$$
where $L_{i(\bphi)}(y_i,\bphi)=\partial L_{i}(y_i,\bphi)/\partial \bphi$ for $L_i(y_i,\bphi)=\log\{p f_1(y_i;\bphi)+(1-p)f_2(y_i;\bphi)\}$.
\end{lem}

\medskip
Note that the conditional asymptotic variance of $\bphih$ does not depend on $y_i$, while the conditional asymptotic bias depends on $y_i$.
Using Lemma \ref{lem:asymp}, we can evaluate the second term as
\begin{align*}
\E\left[(\muh_i-\mut_i)^2|y_i\right]&=\E\left[\left\{\mut_{i(\bphi)}^t(\bphih-\bphi)\right\}^2\Big|y_i\right]+o_p(m^{-1})\\
&=\tr\left(\bOm\mut_{i(\bphi)}\mut_{i(\bphi)}^t\right)+o_p(m^{-1}),
\end{align*}
where $\mut_{i(\bphi)}=\partial \mut_{i}/\partial \bphi$.
Let
\begin{equation}\label{R2}
R_{2i}(y_i,\bphi)\equiv\tr\left(\bOm\mut_{i(\bphi)}\mut_{i(\bphi)}^t\right).
\end{equation}
Since $R_{2i}(y_i,\bphi)=O_p(m^{-1})$, we obtain the following theorem.

% THEOREM ---MSE---
\begin{thm}\label{thm:mse}
Let ${\rm CM}_i^{\ast}(y_i,\bphi)=R_{1i}(y_i,\bphi)+R_{2i}(y_i,\bphi)$ for $R_{1i}$ and $R_{2i}$ given in $(\ref{R1})$ and $(\ref{R2})$, respectively.
Under Assumption 1, we have
$$
{\rm CM}_i(y_i,\bphi)={\rm CM}_i^{\ast}(y_i,\bphi)+o_p(m^{-1}).
$$
\end{thm}

%-------------------------------------------------------------------------------------------------------------------------------
%    MSE ESTIMATION
%-------------------------------------------------------------------------------------------------------------------------------
\subsection{Second-order unbiased estimator of CMSE}\label{sec:estmse}

The approximated CMSE given in Theorem \ref{thm:mse} depends on the unknown parameter $\bphi$, so that it is not feasible in practice.
Here we provide a second-order unbiased estimator of the CMSE.
In what follows, we use the abbreviated notations $R_{1i}$ and $R_{2i}$ instead of $R_{1i}(y_i,\bphi)$ and $R_{2i}(y_i,\bphi)$, respectively, without any confusion.

Since $R_{2i}=O_p(m^{-1})$, we estimate it by the plug-in estimator $R_{2i}(y_i,\bphih)$ with second-order unbiasedness, that is, $\E[R_{2i}(y_i,\bphih)-R_{2i}(y_i,\bphi)|y_i]=o_p(m^{-1})$.
On the other hand, the plug-in estimator $R_{1i}(y_i,\bphih)$ has a second-order bias, namely $\E[R_{1i}(y_i,\bphih)-R_{1i}(y_i,\bphi)|y_i]=O_p(m^{-1})$, because $R_{1i}=O_p(1)$.
To achieve the second-order accuracy, we correct the second-order bias of the estimator $R_{1i}(y_i,\bphih)$.
Using the Taylor series expansion, we have
$$
R_{1i}(y_i,\bphih)=R_{1i}+R_{1i(\bphi)}^t(\bphih-\bphi)+\frac12(\bphih-\bphi)^tR_{1i(\bphi\bphi)}(\bphih-\bphi)+o_p(m^{-1}),
$$
where $R_{1i(\bphi)}=\partial R_{1i}/\partial \bphi$ and $R_{1i(\bphi\bphi)}=\partial^2 R_{1i}/\partial \bphi\partial\bphi^t$.
From Lemma \ref{lem:asymp}, it is seen that the second-order bias in $R_{1i}(y_i,\bphih)$ is 
\begin{align*}
b_i(y_i,\bphi)\equiv &R_{1i(\bphi)}^tE\left(\bphih-\bphi|y_i\right)+\frac12\tr\left(R_{1i(\bphi\bphi)}\E\left[(\bphih-\bphi)(\bphih-\bphi)^t|y_i\right]\right)\\
&=R_{1i(\bphi)}^t\left(\B-\bOm L_{i(\bphi)}\right)+\frac12\tr\left(R_{1i(\bphi\bphi)}\bOm\right).
\end{align*}
Thus, the bias-corrected estimator of $R_{1i}$ is given by
\begin{equation}\label{R1.est}
R_{1i}^{BC}(y_i,\bphih)=R_{1i}(y_i,\bphih)-b_i(y_i,\bphih),
\end{equation}
which satisfies $\E[R_{1i}^{BC}(y_i,\bphih)-R_{1i}(y_i,\bphi)|y_i]=o_p(m^{-1})$.

% THEOREM ---MSE ESTIMATION---
\begin{thm}\label{thm:estmse}
Let $\widehat{{\rm CM}}_i=R_{1i}^{BC}(y_i,\bphih)+R_{2i}(y_i,\bphih)$, where $R_{1i}^{BC}(y_i,\bphih)$ is given in $(\ref{R1.est})$.
Then, under Assumption 1, we have
$$
\E\left[\widehat{{\rm CM}}_i-{\rm CM}_i|y_i\right]=o_p(m^{-1}).
$$
\end{thm}

\medskip
To calculate the $\widehat{{\rm CM}}_i$, we compute the estimates of $\bOm$ and $\B$ using the parametric bootstrap method.
Let $\bOmh$ and $\Bh$ be bootstrap estimators of $\bOm$ and $\B$, respectively.
Then, we have the approximations 
$$
E[\bOmh]=\bOm+o(m^{-1}), \ \ \ \ E[\Bh]=\B+o(m^{-1}),
$$
because $\bOm=O(m^{-1})$ and $\B=O(m^{-1})$.
Moreover, we need to compute $f_{1(\bphi)}$, $f_{2(\bphi)}$, $R_{1i(\bphi)}$, $R_{1i(\bphi\bphi)}$ in $b_i$ and $\mut_{i(\bphi)}$ in $R_{2i}$ at $\bphi=\bphih$.
However, their analytical expressions are too complicated to use them in practice.
Thus we utilize the numerical derivatives which were suggested in Lahiri, Maiti, Katzoff and Parsons (2007).
Let $\{z_m\}$ be a sequence of positive real numbers converging to $0$.
Based on $\{ z_m\}$, we define 
\begin{align*}
&f_{a(\phi_j)}^{\ast}(y_i,\bphih)=\frac{1}{2z_m}\left\{f_a(y_i,\bphih+z_m\e_j)-f_a(y_i,\bphih-z_m\e_j)\right\}, \ \ \ a=1,2\\
&\mut_{i(\phi_j)}^{\ast}(y_i,\bphih)=\frac{1}{2z_m}\left\{\mut_i(y_i,\bphih+z_m\e_j)-\mut_i(y_i,\bphih-z_m\e_j)\right\}\\
&R_{1i(\phi_j)}^{\ast}(y_i,\bphih)=\frac{1}{2z_m}\left\{R_{1i}(y_i,\bphih+z_m\e_j)-R_{1i}(y_i,\bphih-z_m\e_j)\right\}
\end{align*}
where $\e_j$ is a vector of $0$'s other than the $j$-th element is $1$.
Similarly, we define approximations of the second-order partial derivatives of $R_{1i}$ as
\begin{align*}
R_{1i(\phi_j\phi_j)}^{\ast}(y_i,\bphih)=&\frac{1}{z_m^2}\left\{R_{1i}(y_i,\bphih+z_m\e_j)+R_{1i}(y_i,\bphih-z_m\e_j)-2R_{1i}(y_i,\bphih)\right\}, \ \ \ j=1,\ldots,k\\
R_{1i(\phi_j\phi_\ell)}^{\ast}(y_i,\bphih)=&\frac{1}{2z_m^2}\Big[\left\{R_{1i}(y_i,\bphih+z_m\e_{j\ell})+R_{1i}(y_i,\bphih-z_m\e_{j\ell})-2R_{1i}(y_i,\bphih)\right\}\\
&\hspace{1cm} -z_m^2\left\{R_{1i(\phi_j\phi_j)}^{\ast}(y_i,\bphih)+R_{1i(\phi_\ell\phi_\ell)}^{\ast}(y_i,\bphih)\right\}\Big], \ \ \ j\neq \ell,
\end{align*}
where $\e_{j\ell}=\e_j+\e_\ell$.
The justification of the approximations based on these numerical derivatives is given in the following theorem, where the proof is given in the Appendix.

\begin{thm}\label{thm:numderiv}
Under Assumption \ref{as}, we have
\begin{align*}
&|f_{a(\phi_j)}^{\ast}(y_i,\bphih)-f_{a(\phi_j)}(y_i,\bphih)|=O_p(z_m), \ \ \ \ \  |\mut_{i(\phi_j)}^{\ast}(y_i,\bphih)-\mut_{i(\phi_j)}(y_i,\bphih)|=O_p(z_m)\\
&|R_{1i(\phi_j)}^{\ast}(y_i,\bphih)-R_{1i(\phi_j)}(y_i,\bphih)|=O_p(z_m), \ \ \ |R_{1i(\phi_j\phi_\ell)}^{\ast}(y_i,\bphih)-R_{1i(\phi_j\phi_\ell)}(y_i,\bphih)|=O_p(z_m)
\end{align*}
\end{thm}

From Theorem \ref{thm:numderiv}, the second-order unbiasedness of the MSE estimator given in Theorem \ref{thm:estmse} is still valid as far as $z_m=o(m^{-1})$.
In our numerical investigation given in the next section, we use $z_m=m^{-5/4}$.

%-------------------------------------------------------------------------------------------------------------------------------
%    SECTION ---Simulation studies---
%-------------------------------------------------------------------------------------------------------------------------------
\section{Simulation Studies}\label{sec:sim}

%-------------------------------------------------------------------------------------------------------------------------------
%    MODEL BASED SIMULATION
%-------------------------------------------------------------------------------------------------------------------------------
\subsection{Prediction error comparison}
We first evaluated a finite sample performance of the proposed empirical uncertain Bayes method.
Specifically, we compared the EUB estimator with the traditional empirical Bayes (EB) estimator.
We focused on the two models: Poisson-gamma and binomial-beta models described in Section \ref{sec:example}.

For the Poisson-gamma model, we considered the following data generating process:
$$
{\rm PG:} \ \ \ 
(n_iy_i)|\th_i\sim {\rm Po}(n_i\th_i), \ \ \ \th_i|(s_i=1)\sim {\rm Ga}(\nu\exp(\beta_0+\beta_1x_i),\nu), \ \ \ \ P(s_i=1)=p,
$$
where $\beta_0=0, \beta_1=0.5$, $\nu=5$, $m=50$, $n_i$'s were generated from the uniform distribution on $\{5,6,\ldots,30\}$, and $x_i$'s were generated from a standard normal distribution.
The prior probability $p$ takes the values $0.2,0.4,0.6,0.8$ and $1$, where the conventional Poisson-gamma model corresponds to  the data generating process with $p=1$.
We computed both the EUB and EB estimators from the simulated data set.
Based on $R=5,000$ iterations of the data generation, we calculated the mean squared error and the absolute bias which are respectively defined as 
\begin{equation}\label{criteria}
{\rm MSE}_i=\frac{1}{R}\sum_{r=1}^R\left(\thh_i^{(r)}-\th_i^{(r)}\right)^2, \ \ \ \ 
{\rm Bias}_i=\frac{1}{R}\Big|\sum_{r=1}^R\left(\thh_i^{(r)}-\th_i^{(r)}\right)\Big|.
\end{equation}

Define ${\rm MSE}_i({\rm EUB})$ and ${\rm MSE}_i({\rm EB})$ be the simulated values ${\rm MSE}_i$ for the EUB estimator and the EB estimator, respectively.
Then we computed the ratio ${\rm Ra}_i={\rm MSE}_i({\rm EUB})/{\rm MSE}_i({\rm EB})$ for each $i$, and calculated the $q\%$ quantiles of $\{{\rm Ra}_1,\ldots,{\rm Ra}_m\}$ for $q=5,25,50,75$ and $95$.
Hence, if ${\rm Ra}_i$ is smaller than $1$, the EUB estimator performs better than the EB estimator in terms of MSE.
We similarly define the ratio of the absolute biases, and the results for the five $p$ patterns are given in Table \ref{tab:comp}.
In the scenario $p=1$, the traditional Poisson-gamma model is the true model and uncertain model is overfitting.
However, the results show that the EUB estimator performs as well as the EB estimator, which indicates that the effect of overfitting seems small.
Moreover, from Table \ref{tab:comp}, when $p$ is smaller than $1$,  it is revealed that the EUB estimator improve the EB estimator in terms of both the MSE and the absolute bias, and the improvement is greater as $p$ gets smaller.

We next compared performances of the two estimators in the binomial-beta model using the data generating process:
$$
{\rm BB:} \ \ \ 
(n_iy_i)|\th_i\sim {\rm Bin}(n_i,\th_i), \ \ \ \th_i|(s_i=1)\sim {\rm Beta}(\nu m_i,\nu(1-m_i)), \ \ \ \ P(s_i=1)=p,
$$
with $m_i=\exp(\beta_0+\beta_1x_i)/\{1+\exp(\beta_0+\beta_1x_i)\}$, where $\beta_0=0, \beta_1=0.5$, $\nu=5$, $m=50$, $n_i$'s were generated from the uniform distribution on $\{10,11,\ldots,30\}$, and $x_i$'s were generated from a standard normal distribution.
Similarly to the previous study, we simulated the MSE and the absolute bias using (\ref{criteria}) with $R=5,000$, and computed the quantiles of the ratios.
The results are given in Table \ref{tab:comp}, which shows the similar results to the Poisson-gamma case.
However, the amount of improvement seems smaller than that in the Poisson-gamma case, but the EUB estimator performs better than the EB estimator in the binomial-beta case.

% Table
\begin{table}[htb!]
\caption{Simulated ratios of the MSEs and absolute biases of the EUB estimator over the EB estimator.
\label{tab:comp}}
\begin{center}
\begin{tabular}{cccccccccccccc}
\toprule
& && \multicolumn{5}{c}{MSE} & &  \multicolumn{5}{c}{Absolute bias}\\ 
& $p$ && 5\% & 25\% & 50\% & 75\% & 95\% & &  5\% & 25\% & 50\% & 75\% & 95\%\\
\midrule
 & 0.2 &  & 0.156 & 0.355 & 0.542 & 0.717 & 0.842 &  & 0.554 & 0.628 & 0.707 & 0.832 & 0.878 \\
 & 0.4 &  & 0.298 & 0.454 & 0.578 & 0.710 & 0.815 &  & 0.652 & 0.693 & 0.726 & 0.799 & 0.874 \\
PG & 0.6 &  & 0.540 & 0.640 & 0.728 & 0.826 & 0.894 &  & 0.795 & 0.821 & 0.843 & 0.876 & 0.911 \\
 & 0.8 &  & 0.700 & 0.832 & 0.876 & 0.912 & 0.962 &  & 0.890 & 0.911 & 0.925 & 0.949 & 0.977 \\
 & 1 &  & 0.985 & 0.993 & 1.000 & 1.005 & 1.011 &  & 0.989 & 0.993 & 0.999 & 1.002 & 1.007 \\
 \midrule
 & 0.2 &  & 0.395 & 0.608 & 0.730 & 0.807 & 0.996 &  & 0.490 & 0.583 & 0.650 & 0.768 & 0.980 \\
 & 0.4 &  & 0.488 & 0.736 & 0.827 & 0.887 & 0.983 &  & 0.721 & 0.747 & 0.805 & 0.862 & 0.987 \\
BB & 0.6 &  & 0.730 & 0.866 & 0.909 & 0.934 & 0.983 &  & 0.813 & 0.851 & 0.886 & 0.924 & 0.994 \\
 & 0.8 &  & 0.865 & 0.946 & 0.970 & 0.984 & 0.993 &  & 0.916 & 0.938 & 0.961 & 0.971 & 0.992 \\
 & 1 &  & 0.966 & 0.980 & 1.001 & 1.007 & 1.017 &  & 0.979 & 0.987 & 0.995 & 1.004 & 1.012\\
\bottomrule
\end{tabular}
\end{center}
\end{table}

%-------------------------------------------------------------------------------------------------------------------------------
%    Sensitivity 
%-------------------------------------------------------------------------------------------------------------------------------
\subsection{Sensitivity to distributional assumptions}
We next investigated sensitivity to distributional assumptions in the proposed model.
Here we focused on the Poisson-gamma model as considered in the previous simulation study:
$$
(n_iy_i)|\th_i\sim {\rm Po}(n_i\th_i), \ \ \ \th_i|(s_i=1)\sim {\rm Ga}(\nu\exp(\beta_0+\beta_1x_i),\nu), \ \ \ \ P(s_i=1)=p,
$$
where $p=0.5$, and other settings $\beta$, $\beta_1$, $\nu$, $n_i$ and $x_i$ are set as the same values as in the previous section. 
To asses sensitivity of the distributional assumption of the proposed method, we consider the two alternative distributions: a log-normal distribution and a two-point distribution for $\th_i$ instead of the gamma distribution.
Noting that $\E[\th_i]=m_i$ and $\Var(\th_i)=m_i/\nu$ under the gamma distribution, we scaled two distributions to have the same expectation and variance.
Specifically, we set $\log \th_i\sim N(\log(m_i/\sqrt{1+1/\nu m_i}),\log(1+1/\nu m_i))$ for the log-normal distribution, and $P(\th_i=m_i+\sqrt{m_i/\nu})=P(\th_i=m_i-\sqrt{m_i/\nu})=0.5$ for the two-point distribution.
Based on $R=5,000$ simulation runs, we computed the MSE and absolute bias with the formula (\ref{criteria}) for three underlying distributions.
In Table \ref{tab:sens}, we show the five quantiles of the simulated MSE and absolute bias.
It is observed that both the MSE and the absolute bias in the misspecified cases of log-normal and two-point distributions are larger than the correctly specified case of the gamma distribution.
The absolute biases in the two misspecified cases are about twice as large as that in the gamma case, so that the inflation of the absolute bias seems relatively large.
However, the difference in the MSE is around $10\%$, so that the effect of misspecification on MSE seems relatively small.

% Table
\begin{table}[htb!]
\caption{Quantiles of the simulated MSE and absolute bias of the EUB estimator under the three underlying distributions (the values are multiplied by 100).
\label{tab:sens}}
\begin{center}
\begin{tabular}{cccccccccccccc}
\toprule
 && \multicolumn{5}{c}{MSE} & &  \multicolumn{5}{c}{Absolute bias}\\ 
distribution && 5\% & 25\% & 50\% & 75\% & 95\% & &  5\% & 25\% & 50\% & 75\% & 95\%\\
\midrule
Gamma &  & 1.31 & 2.55 & 3.98 & 5.92 & 11.06 &  & 0.03 & 0.10 & 0.25 & 0.48 & 0.71 \\
Log-normal &  & 1.32 & 2.47 & 4.15 & 5.85 & 10.89 &  & 0.05 & 0.33 & 0.49 & 0.72 & 1.25 \\
two-point &  & 1.57 & 2.87 & 4.50 & 6.46 & 10.58 &  & 0.08 & 0.28 & 0.51 & 0.84 & 1.18\\
\bottomrule
\end{tabular}
\end{center}
\end{table}

%-------------------------------------------------------------------------------------------------------------------------------
%    Performances of CMSE estimator 
%-------------------------------------------------------------------------------------------------------------------------------
\subsection{Finite sample performance of the CMSE estimator}
Finally we investigated a finite sample behavior of the CMSE estimator provided in Theorem \ref{thm:estmse} in the UPG model.
We considered the simple data generating process without covariates:
\begin{equation}\label{PG}
(n_iy_i)|\la_i\sim {\rm Po}(n_i\la_i), \ \ \ \la_i|(s_i=1)\sim {\rm Ga}(\nu\exp(\beta),\nu), \ \ \ \ P(s_i=1)=p,
\end{equation}
with $\beta=1$, $\nu=5$, $p=0.5$ and $n_i=10$.
For the number of areas,  we consider the two cases of $m=50$ and $m=100$.
For conditioning values of $y_{\alpha}$, we consider $\alpha$-quantiles of the marginal distribution of $y_i$, where $\alpha=0.1,0.2,\ldots,0.9$, and calculate these values by generating $10,000$ random samples from (\ref{PG}).
To get the simulated values of the CMSE given $y_{\alpha}$, we generate random samples from (\ref{PG}) and replace $y_1$ with $y_{\alpha}$, and we computed the EUB estimator of $\lah_1$.
For the true values of $\la_1$, since $y_{\alpha}$ is given, we generate $\la_1$ from the posterior distribution $\la_1|y_{\alpha}\sim r_1{\rm Ga}(\lat_1,(n_1+\nu)^{-1}\lat_1)+(1-r_1)\de_{\la_1}(\exp(\beta))$ with $\lat_1=(n_1+\nu)^{-1}(n_1y_{\alpha}+\nu\exp(\beta))$ and $r_1$ given in (\ref{PG-prob}).
Then, based on $R=10,000$ iteration, we calculate the simulated values of the CMSE defined as
\begin{equation*}
{\rm CM}_{\alpha}=\frac1R\sum_{r=1}^R(\lah_1^{(r)}-\la_1^{(r)})^2, 
\end{equation*}
where $\lah_1^{(r)}$ and $\la_1^{(r)}$ are the EUB estimates of $\la_1$, and $\la_1^{(r)}$ is the generated value from the distribution of $\la_1|y_{\alpha}$ in the $r$th iteration.

For evaluation of the CMSE estimator, we generated random samples from (\ref{PG}) and replace $y_1$ with $y_{\alpha}$, and get CMSE estimators with $B=100$ bootstrap samples and $z_m=m^{-5/4}$ for computing the numerical derivatives.
This procedure is repeated $S=2,000$ times and calculated the percentage relative bias (RB) and the coefficient of variation (CV) defined as 
$$
{\rm RB}_{\alpha}=\frac{1}{S}\sum_{s=1}^{S}\left(\frac{{\widehat {\rm CM}}_{\alpha}-{\rm CM}_{\alpha}}{{\rm CM}_{\alpha}}\right)\times 100, \ \ \ \ {\rm CV}_{\alpha}=\sqrt{\frac{1}{S}\sum_{s=1}^{S}\left(\frac{{\widehat {\rm CM}}_{\alpha}-{\rm CM}_{\alpha}}{{\rm CM}_{\alpha}}\right)^2}.
$$
Remember that the suggested CMSE estimator given in Theorem \ref{thm:estmse} is second-order unbiased.
To emphasize the importance of bias correction in estimating the CMSE, we also computed the two criteria of the naive CMSE estimator defined as ${\widehat {\rm CM}}_{\alpha (N)}=R_{11}(y_{\alpha},\bphih)$.
It is noted that the naive estimator has the first order bias, because it ignores the second term $R_{2i}$ and the bias of the plug-in estimator of $R_{1i}$.
For the naive estimator, we calculated RB and CV based on the same number of iteration and we define RBN and CVN as RB and CV of the naive estimator.
The resulting values are given in Table \ref{tab:cmse} for both $m=50$ and $m=100$.

It is observed from Table \ref{tab:cmse}, the naive estimator has the serious negative bias when $m=50$. 
Especially, when the condition values are upper or lower quantiles, the negative bias tends to be larger.
This comes from the fact that the naive estimator ignores the positive $O_p(m^{-1})$ term in the CMSE decomposition given in Theorem \ref{thm:mse}.
Since practitioners decide policies or investments based on estimated values as well as their risk estimates, the under-estimation of the CMSE is considered serious in practice.
Hence, the results in Table \ref{tab:cmse} show that the naive CMSE estimator without bias correction is not suitable for practical use.
On the other hand, the bias-corrected CMSE estimator works well in both $m=50$ and $m=100$ and provides accurate estimation of the CMSE in terms of the relative biases.
Concerning the CV values, the bias-corrected estimator has a slightly larger CV than the naive estimator in most cases.
This is because the bias corrected terms increase the variance of the estimator.
However, the difference is not so significant.
Thus, the bias-corrected CMSE estimator is useful in practice.

% table
\begin{table}[htb!]
\caption{Percentage of relative bias and coefficient of variation.
\label{tab:cmse}}
\begin{center}
\begin{tabular}{cccccccccccccc}
\toprule
&$\alpha$&& 0.1 & 0.2 & 0.3 & 0.4 & 0.5 & 0.6 & 0.7 & 0.8 & 0.9\\
&$y_i$ && 1.8 & 2.1 & 2.3 & 2.5 & 2.7 & 2.9 & 3.0 & 3.3 & 3.6\\
 \midrule
$m=50$ &RB && 1.21 & -1.01 &  6.87 &  6.56 &  5.87 &  3.43 &  2.42 &-3.95 & -9.59\\
&RBN &&-27.7 &-23.5 & -13.8 &  -6.56 &  -3.28 &  -3.86 & -8.37 & -21.1 & -28.6\\
&CV && 0.53 & 0.36 & 0.53 & 0.67 & 0.73 & 0.68 & 0.64 &0.43 & 0.34\\
&CVN && 0.38 & 0.37 & 0.43 & 0.53 & 0.61 & 0.55 & 0.51& 0.39 & 0.39\\
\midrule
$m=100$ &RB && -0.23 & -0.45  & 3.28 &  3.68 &  2.02 & -0.12 &  0.55 & -2.94 & -5.82\\
&RBN &&   -16.9 & -12.9 &  -6.33 &   -3.75   &4.79 &   0.97 & -2.55 &-12.4 &-17.9\\
&CV &&  0.28 & 0.25 & 0.41 & 0.57 & 0.63 & 0.58 & 0.52 & 0.32 & 0.24\\
&CVN && 0.28 & 0.27 & 0.35 & 0.49 & 0.52& 0.49 & 0.42 & 0.29 & 0.27\\
\bottomrule
\end{tabular}
\end{center}
\end{table}

%-------------------------------------------------------------------------------------------------------------------------------
%    ---Section---  EMPIRICAL APPLICATION
%-------------------------------------------------------------------------------------------------------------------------------
\section{Illustrative Examples}\label{sec:application}

%-------------------------------------------------------------------------------------------------------------------------------
%    Mortality data
%-------------------------------------------------------------------------------------------------------------------------------
\subsection{Historical mortality data in Tokyo}
The mortality rate is a representative index in demographics and has been used in various fields.
Especially, in economic history, one can discover new knowledge from a spatial distribution of mortality rate in small areas.  
As divisions get smaller (e.g. city$\to$town$\to$block$\ldots$), one can get a more informative spatial distribution.
However, the direct estimate of the mortality rate in small area with extremely low population has high variability, which may leads to incorrect recognition of the spatial distribution.
Therefore, it is desirable to use smoothed and stabilized estimates through empirical Bayes methods.

We here focus on the mortality data in Tokyo, 1930.
The data set consists of the observed mortalities $z_i$ and the number of population $N_i$ in the $i$th area in Tokyo.
Such area-level data are available for $m=1,371$ small areas.
We first computed the expected mortality in the $i$th area as $n_i=N_i\sum_{j=1}^m z_j/\sum_{j=1}^mN_j$.
The standardized mortality ratio (SMR) is defined as the ratio of the actual mortality to the expected mortality for each area, which is often used in epidemiology as an indicator of potential mortality risk (Rothman, Greenland and Lash, 2008).
Then, the direct estimator of the SMR in the $i$th area is $y_i=z_i/n_i$.
It is noted that $y_i=0$ in $84$ areas, the number of areas with SMR larger than $1$ is $526$, and the maximum value of $y_i$ is $16.4$.

\begin{figure}[htbp!]
\begin{center}
\includegraphics[width=7.3cm]{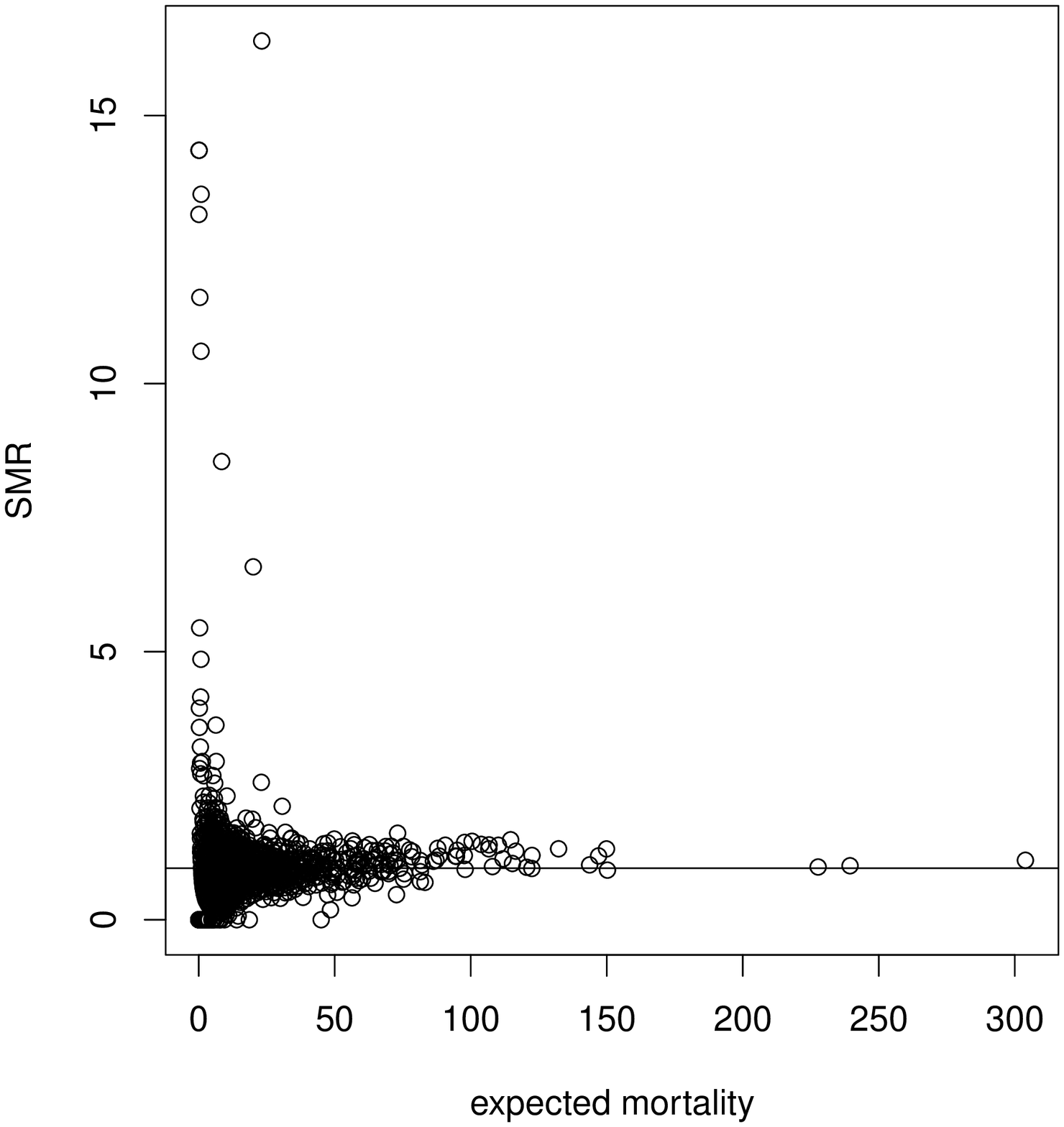}\ \ 
\includegraphics[width=7.3cm]{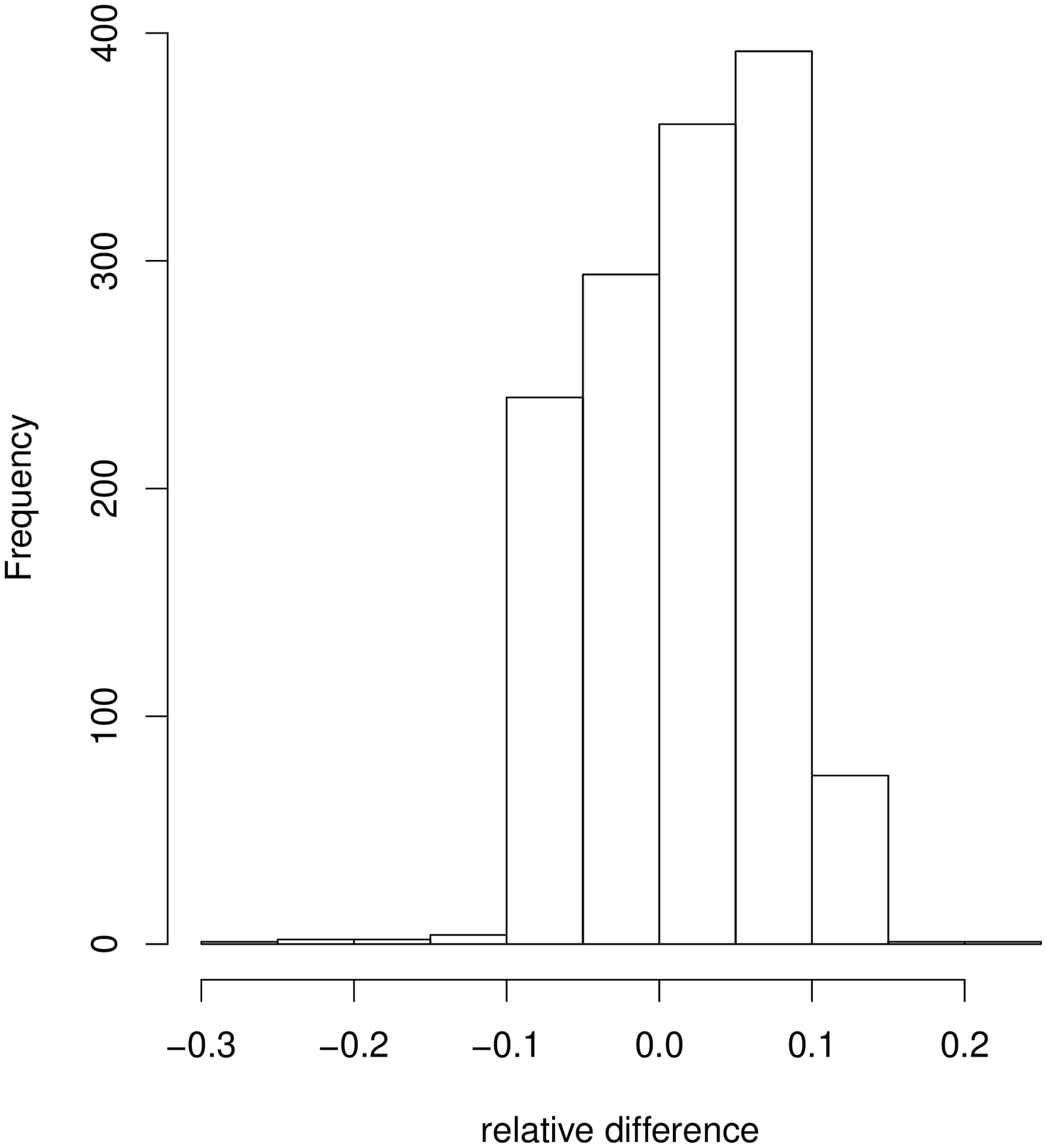}\\
\includegraphics[width=7.3cm]{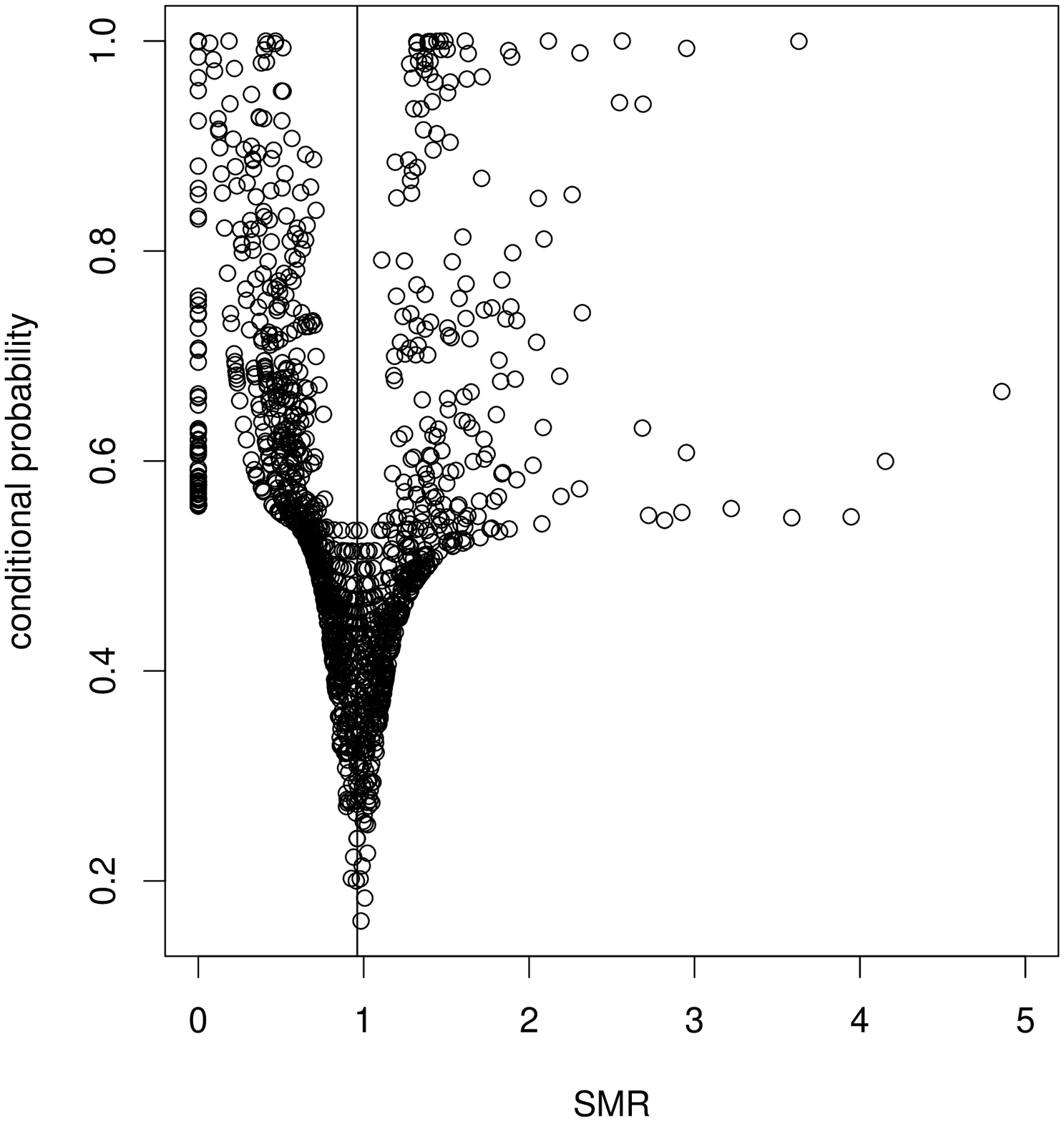}\ \ 
\includegraphics[width=7.3cm]{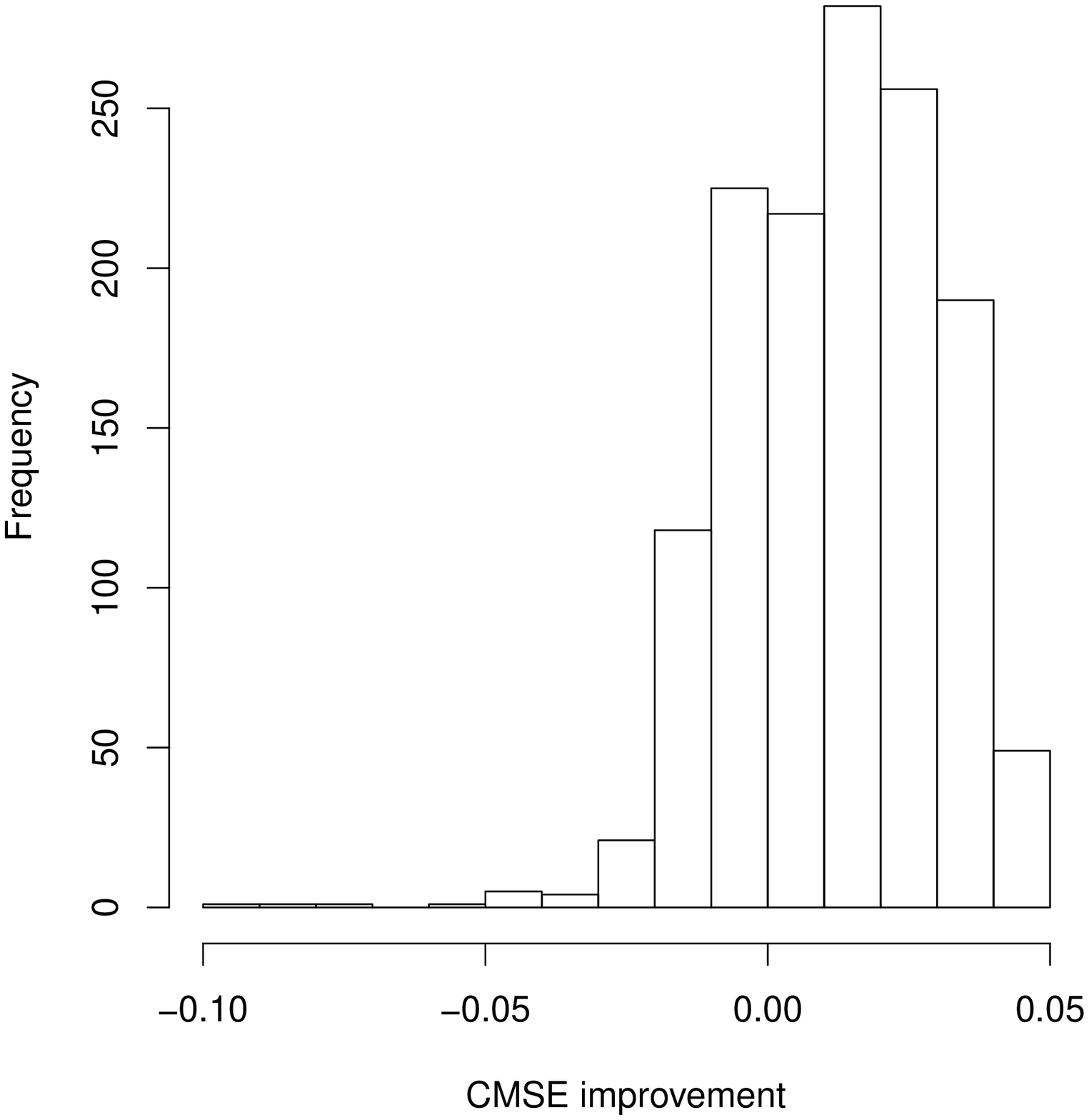}
\caption{Sample plot of the expected mortality $n_i$ and the SMR $y_i$ (upper-left), the scaled difference of two predictors: $(\lah_i^{\rm UPG}-\lah_i^{\rm PG})/\lah_i^{\rm PG}$ (upper-right), sample plot of the estimates of conditional probability $P(s_i=1|y_i)$ and the SMR $y_i$ (lower-left), and histogram of improvement of estimated CMSE: $\widehat{\rm  CM}^{\rm PG}_i-\widehat{\rm  CM}^{\rm UPG}_i$ (lower-right).
\label{fig:SMR}
}
\end{center}
\end{figure}

For this data set, we apply the two models: the uncertain Poisson-gamma model described Section \ref{sec:example} and the traditional Poisson-gamma model, described as 
\begin{align*}
{\rm UPG:}\ \ &
n_iy_i|\la_i\sim {\rm Po}(n_i\la_i), \ \ \ \la_i|s_i\sim {\rm Ga}(\nu\exp(\beta),\nu), \ \ \ \ P(s_i=1)=p\\
{\rm PG:}\ \ &
n_iy_i|\la_i\sim {\rm Po}(n_i\la_i), \ \ \ \la_i\sim {\rm Ga}(\nu\exp(\beta),\nu), 
\end{align*}
where $\la_i=E(y_i|\la_i)$ denotes the `true' SMR in the $i$-th area, which we want to estimate.
Using the EM algorithm in Section \ref{sec:EM} with $5,000$ Monte Calro samples in each E-step, we get the point estimates of the parameters of the two models as shown in Table \ref{tab:SMR}.
For comparison of the two models, we computed AIC and BIC based on the maximum marginal likelihood, and the results are also given in Table \ref{tab:SMR}.
In terms of AIC and BIC, the proposed UPG model fits better than the traditional PG model for this data set.
This comes from the feature of the data.
In the upper left panel of Figure \ref{fig:SMR}, we show the sample plot of the expected mortality $n_i$ and the SMR $y_i$, noting that the solid line corresponds to the estimated regression line $y_i=\exp(\beh)$ in the UPG model.
It is observed that most $y_i$ are distributed around the regression line, and the random area effects are necessary in most areas.
The UPG model tells us about the feature of the data through the estimate of $p$.
The lower left panel of Figure \ref{fig:SMR} provides a scatter plot of the estimated conditional probability $P(s_i=1|y_i)$ and the SMR $y_i$, where the conditional probability $P(s_i=1|y_i)$ corresponds to the probability of existing random area effect in the $i$th area when $y_i$ is observed.
The solid line corresponds to the estimated regression line $y_i=\exp(\beh)$ in the UPG model.
From the figure, we can see that the estimates of $P(s_i=1|y_i)$ are dramatically different from area to area, and the probability gets lower as SMR is closer to the regression line.
To see the difference of estimated values of $\la_i$, in the upper right panel of Figure \ref{fig:SMR}, we present the relative differences between estimators from the two models, which are defined as $(\lah_i^{\rm UPG}-\lah_i^{\rm PG})/\lah_i^{\rm PG}$, where $\lah_i^{\rm UPG}$ and $\lah_i^{\rm PG}$ are empirical Bayes estimates of $\la_i$ from the UPG model and the PG model, respectively.
We can observe that the differences are around $10\%$ and are not negligible.

We next calculated the CMSE estimates of the EUB estimates $\lah_i^{\rm UPG}$ using Theorem \ref{thm:estmse} with $B=100$ and $z_m=m^{-5/4}$.
For comparison, we also computed the CMSE estimates of $\lah_i^{\rm PG}$ using Theorem \ref{thm:estmse} with $p=1$, $B=100$ and $z_m=m^{-5/4}$.
Then, we computed their difference and their histogram over areas is given in the lower right panel of Figure \ref{fig:SMR}.
In the figure, the positive value indicates that the EUB estimator has the smaller CMSE value than the EB estimator, and it is revealed that the EUB estimator can improve the estimation risk over the EB estimator in many areas.
In particular, the mean values of CMSEs are $4.2\times 10^{-2}$ for UPG and $5.4\times 10^{-2}$ for PG, so that the EUB estimator can improve $20\%$ CMSE values over the traditional EB estimator on average.

Finally, we assessed the performance of the two models in terms of prediction accuracy in non-sampled areas.
Since areas with small $n_i$ have high variability, we consider to predict $y_i$ of areas with $n_i$ larger than the $\alpha$-quantile of $n_i$'s, denoted by $q_{\alpha}$.
Thus we omitted areas with $n_i$ larger than $q_{\alpha}$ and computed the estimates of the model parameters using the remaining data.
For fixed $\alpha$, we define the predictive criterion (PC) as
\begin{equation}\label{PC}
{\rm PC}_{\alpha}=\sum_{i=1}^mI(n_i>q_{\alpha})\left(\mh_i-y_i\right)^2\bigg/\sum_{i=1}^m I(n_i>q_{\alpha}),
\end{equation}
noting that $\mh_i$ is the best predictor in non-sampled areas.
In this example, $\mh_i=\exp(\beh)$.
The values of PC were computed for three quantiles of $\alpha=0.90, 0.95$ and $0.99$ and reported in Table \ref{tab:SMR}.
It is revealed that the EUB method can improve PC values over the EB method by about $10\%$.

% table  ---UPG---
\begin{table}[htb!]
\caption{Point estimates of the model parameters and values of AIC, BIC and PC (multiplied by $100$) for the three thresholds.
\label{tab:SMR}}
\begin{center}
\begin{tabular}{cccccccccccccc}
\toprule
Estimates & $\beh$ & $\nuh$ & $\ph$ & AIC & BIC & PC$_{0.90}$ & PC$_{0.95}$ & PC$_{0.99}$\\
\midrule
UPG & -0.039 & 5.15 & 0.56 & 8142.17 & 8157.84 & 8.00 & 7.70 &  5.69  \\
PG & -0.052 & 7.42 & --- & 8265.12 & 8275.57 & 8.67 & 8.24 & 6.15\\
\bottomrule
\end{tabular}
\end{center}
\end{table}

%  figure ---UPG---

%-------------------------------------------------------------------------------------------------------------------------------
%    Poverty estimates
%-------------------------------------------------------------------------------------------------------------------------------
\subsection{Poverty rates in Spanish provinces}
\begin{figure}[htbp!]
\begin{center}
\includegraphics[width=7.3cm]{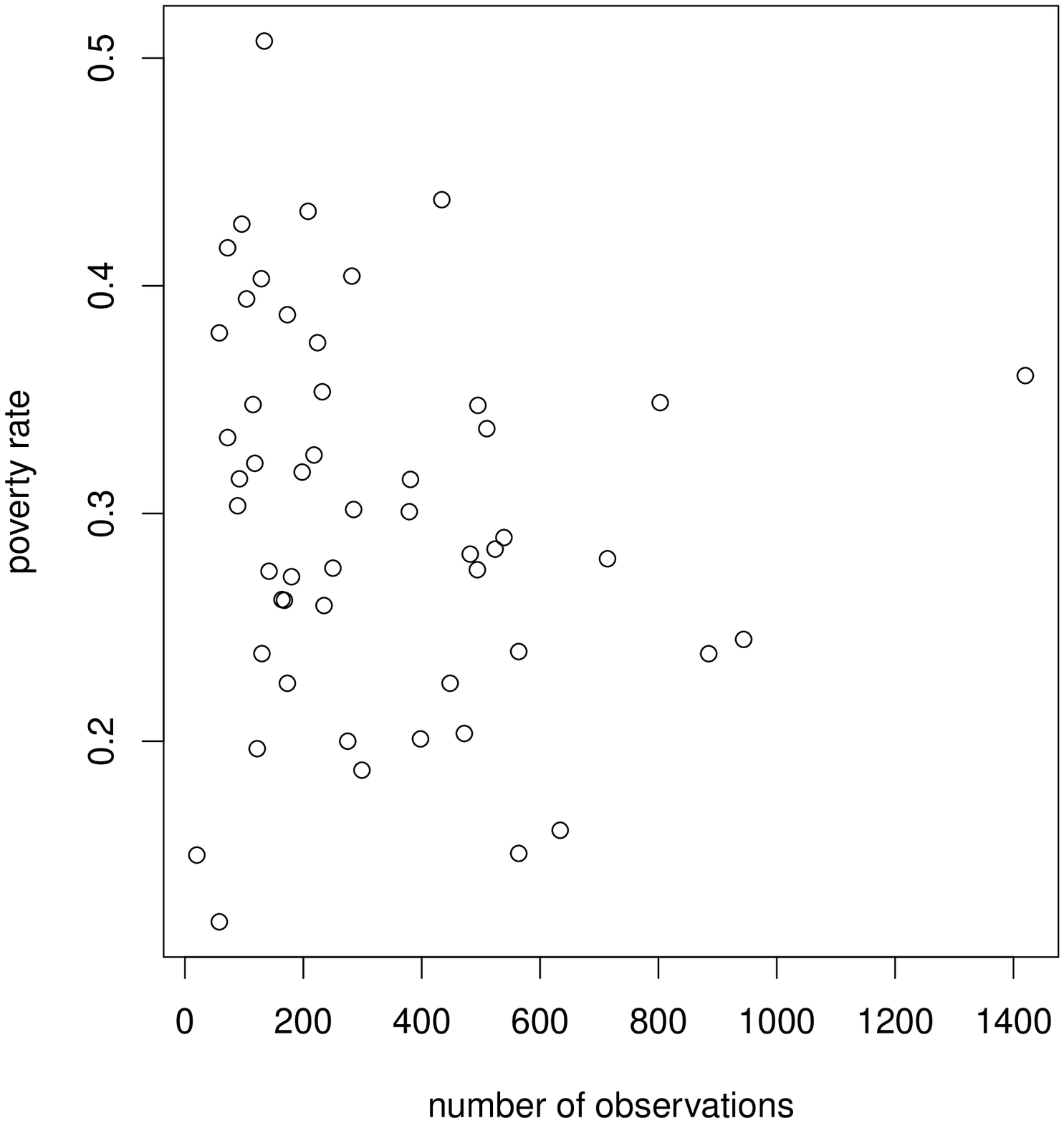}\ \ 
\includegraphics[width=7.3cm]{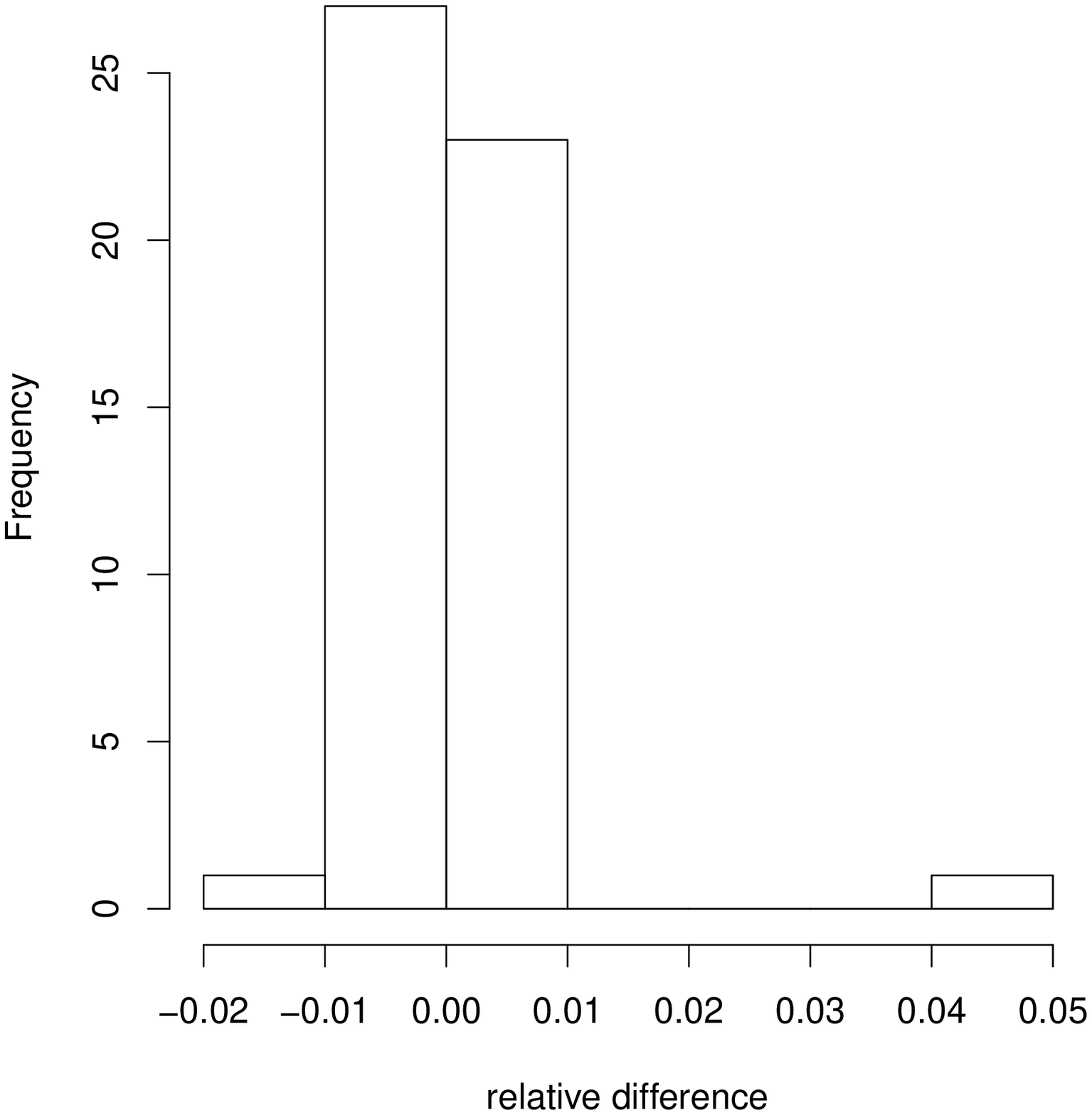}
\caption{
The scatter plot of the number of observations $n_i$ and the direct estimate of poverty rate $y_i$ (left), and the histogram of relative difference $(\thh_i^{\rm UBB}-\thh_i^{\rm BB})/\thh_i^{\rm BB}$ (right).
\label{fig:PR}
}
\end{center}
\end{figure}

We next applied our method to the synthetic income data set in Spanish provinces, which is available in the R package \verb+sae+.
In this application, we focus on estimating area-level poverty rates.
We set the poverty level as $0.7$ times the median of all the observed incomes, and computed the direct estimates of the poverty rates.
As covariates, we calculated area-level rates of female and labors.
The scatter plot of the pairs $(n_i,y_i)$ is given in the left panel of Figure \ref{fig:PR}, from which we can observe that the direct estimate $y_i$ has higher variability as $n_i$ gets smaller.

For the data set, we applied the two models: the uncertain binomial-beta (UBB) model and the traditional binomial-beta (BB) model, described as
\begin{align*}
{\rm UBB:}\ \ &
n_iy_i|\th_i\sim {\rm Bin}(n_i,\th_i), \ \ \ \th_i|s_i\sim {\rm Beta}(\nu m_i,\nu(1-m_i)), \ \ \ \ P(s_i=1)=p\\
{\rm BB:}\ \ &
n_iy_i|\th_i\sim {\rm Bin}(n_i,\th_i), \ \ \ \th_i\sim {\rm Beta}(\nu m_i,\nu(1-m_i)),
\end{align*}
where $y_i$ is the direct estimate of the true poverty rate $p_i$, $n_i$ is the number of observations in the $i$th area, $m_i=\text{logit}(\beta_0+\beta_1g_i+\beta_2f_i)$ for $\text{logit}(x)=\exp(x)/(1+\exp (x))$, and $g_i$ and $f_i$ are rates of populations of females and labors, respectively.
The point estimates of the model parameters based on the EM algorithm in Section \ref{sec:EM} with $5,000$ Monte Carlo samples are shown in Table \ref{tab:PR}.
The signs of $\beh_1$ and $\beh_2$ are reasonable.
From the table, it is observed that the estimate of $p$ in the UBB model is almost $1$, which implies that the traditional BB model is appropriate for this data set.
Actually, the values of AIC and BIC based on the marginal likelihood, given in Table \ref{tab:PR}, support the BB model rather than the UBB model.
Concerning the differences of predicted values, we provide in the right panel of Figure \ref{fig:PR} the histogram of the relative differences: $(\thh_i^{\rm UBB}-\thh_i^{\rm BB})/\thh_i^{\rm BB}$, where $\thh_i^{\rm UBB}$ and $\thh_i^{\rm BB}$ are predicted values from the UBB and the BB models.
It shows that the differences are smaller than $1\%$ in most areas.

We next calculated the CMSE estimates of the EUB and the EB estimates using Theorem \ref{thm:estmse} with $B=100$ and $z_m=m^{-5/4}$.
These two estimates are expected to be similar, but the CMSE estimates of the EUB estimates are negative in some areas, while those of the EB estimates are all positive.
This may comes from the instability of estimating $p$ close to $1$.

Finally, we considered the performances of prediction for non-sampled areas.
Similarly to the previous section, we considered the predictive criterion (PC) defined in (\ref{PC}) with $\mh_i=\text{logit}(\beh_0+\beh_1g_i+\beh_2f_i)$.
The values of PC were computed for $\alpha=0.70, 0.80$ and $0.90$ and reported in Table \ref{tab:PR}.
It is observed that the UBB model provides the performance better than the BB model while the differences are quite small.

% table  ---UBB---
\begin{table}[htb!]
\caption{Point estimates of the model parameters, values of AIC, BIC and PC (multiplied by $1,000$) for the three thresholds.
\label{tab:PR}}
\begin{center}
\begin{tabular}{cccccccccccccc}
\toprule
Estimates & $\beh_0$ & $\beh_1$ & $\beh_2$ & $\nuh$ & $\ph$ & AIC & BIC & PC$_{0.70}$ & PC$_{0.80}$ & PC$_{0.90}$\\
\midrule
UBB & -1.92  & 2.91 & -1.03 & 41.33 & 0.96 & 459.67 & 469.42 &  5.59  & 5.42 & 5.17 \\
BB & -2.14 & 3.36 & -1.07 & 42.93 & --- &  457.74 &  465.55 & 5.61 & 5.42 & 5.19\\
\bottomrule
\end{tabular}
\end{center}
\end{table}

%  figure ---UBB---

%-------------------------------------------------------------------------------------------------------------------------------
%    CONCLUDING
%-------------------------------------------------------------------------------------------------------------------------------
\section{Concluding Remarks}
In this article, we have developed the empirical uncertain Bayes (EUB) methods in area-level models based on the natural exponential family with quadratic variance functions.
We have provided the estimators of the model parameters and the EUB estimator as an alternative of the usual empirical Bayes estimator.
As examples, we consider the three models often used in practice, namely, the Fay-Herriot, Poisson-gamma and binomial-beta models, and discussed shrinkage properties of the EUB estimator.
For the risk evaluation, in the principal of Booth and Hobert (1998), we have constructed the second-order unbiased estimator of the conditional mean squared error.
Through simulation studies and empirical applications, we have confirmed that the proposed EUB estimator performs better than the usual empirical Bayes estimator and the suggested CMSE estimator works well in finite samples.
Since a hierarchical Bayesian approach enables us to obtain more information about the parameter of interest such as a credible interval, it may be a valuable future study.
The data sets and \verb+R+ codes used in Section \ref{sec:application} are available in the supplementary material.

\ \\
{\bf Acknowledgement}

\medskip
We would like to thank the Editor, the Associate Editor and two reviewers for many valuable comments and helpful suggestions which led to an improved version of this paper.
The first author was supported in part by Grant-in-Aid for Scientific Research (15J10076) from Japan Society for the Promotion of Science (JSPS).
Research of the second author was supported in part by Grant-in-Aid for Scientific Research  (15H01943 and 26330036) from Japan Society for the Promotion of Science.
Research of the third author was supported in part by Grant-in-Aid for Scientific Research (16K17153) from Japan Society for the Promotion of Science.

%-------------------------------------------------------------------------------------------------------------------------------
%    APPENDIX
%-------------------------------------------------------------------------------------------------------------------------------
\vspace{0.5cm}
\begin{center}
{\large \bf Appendix}
\end{center}

\vspace{0.4cm}
% A1 ---Assumptions---
\noindent
{\bf A1. Checking (iii) in Assumption \ref{as}.}

\medskip\noindent
{\bf (Fay-Herriot model)}.\ \ \ 
It follows from (\ref{FH:dens}) that
$$
f_{1(\bbe)}(y_i;\bphi)=f_{1}(y_i;\bphi)\left(\frac{y_i-\x_i^t\bbe}{A+D_i}\right)\x_i, \ \ \ 
f_{1(A)}(y_i;\bphi)=\frac{f_{1}(y_i;\bphi)}{2(A+D_i)^2}\left\{\left(y_i-\x_i^t\bbe\right)^2-A-D_i\right\}.
$$
Using $f_{1}(y_i;\bphi)\leq 1/\sqrt{2\pi A}$, we can see that $|f_{1(\phi_j)}(y_i;\bphi)|$, $|f_{1(\phi_j\phi_{\ell})}(y_i;\bphi)|$ and $|f_{1(\phi_j\phi_{\ell}\phi_k)}(y_i;\bphi)|$ can be evaluated from above by  6th order polynomials of $y_i$ and the assumption (iii) is satisfied for $a=1$.
The case of $a=2$ can be shown similarly.

\medskip\noindent
{\bf (Poisson-gamma model)}.\ \ \ 
It is noted that $f_{1(\phi_k)}(y_i;\bphi)=f_1(y_i;\bphi)\partial\log f_1(y_i;\bphi)/\partial\phi_k$.
From (\ref{PG:dens}), it holds that 
\begin{align*}
&\frac{\partial\log f_1(y_i;\bphi)}{\partial\beta_k}
=\nu x_{ik}m_i\left\{\psi(n_iy_i+\nu m_i)-\psi(\nu m_i)\right\}, \ \ k=1,\ldots,q,\\
&\frac{\partial\log f_1(y_i;\bphi)}{\partial\nu}
=m_i\left\{\psi(n_iy_i+\nu m_i)-\psi(\nu m_i)\right\}-\frac{n_iy_i}{n_i+\nu}+m_i\left\{\frac{n_i}{n_i+\nu}+\log\left(\frac{n_i}{n_i+\nu}\right)\right\},
\end{align*}
where $\psi(\cdot)$ is the digamma function $\psi(x)=\text{d}\log \Ga(x)/\text{d}x$.
Using the fact that $\psi(x)\approx \log x$ for large $x$, we have $|\partial\log f_1(y_i;\bphi)/\partial\beta_k|=O_p(\log y_i)$ and $|\partial\log f_1(y_i;\bphi)/\partial\nu|=O_p(y_i)$ for large $y_i$.
Since there exists $c\in\Re$ such that $f_1(y_i;\bphi)\leq c$, $|f_{1(\phi_k)}(y_i;\bphi)|$ is bounded above by an liner function of $y_i$.
Concerning the second derivatives, we note that 
\begin{equation}\label{2nd.deriv}
f_{1(\phi_k\phi_\ell)}(y_i;\bphi)
=f_1(y_i;\bphi)\left\{\frac{\partial^2\log f_1(y_i;\bphi)}{\partial\phi_k\partial\phi_\ell}+\frac{\partial\log f_1(y_i;\bphi)}{\partial\phi_k}\frac{\partial\log f_1(y_i;\bphi)}{\partial\phi_\ell}\right\}.
\end{equation}
Moreover, the straightforward calculation shows that 
$$
\frac{\partial^2\log f_1(y_i;\bphi)}{\partial\beta_k^2}
=\nu x_{ik}^2m_i\left\{\psi(n_iy_i+\nu m_i)-\psi(\nu m_i)\right\}+\nu x_{ik}^2m_i^2\left\{\psi^{(1)}(n_iy_i+\nu m_i)-\psi^{(1)}(\nu m_i)\right\},
$$
where $\psi^{(n)}(x)=\text{d}^n\psi(x)/\text{d}x^n$ is a polygamma function. 
Since $\psi^{(n)}(x)\approx (-1)^n(n-1)!x^{-n}$ for large $x$, we have $|\partial^2\log f_1(y_i;\bphi)/\partial\beta_k^2|=O(\log y_i)$ for large $y_i$.
Similarly, we obtain $|\partial^2\log f_1(y_i;\bphi)/\partial\beta_k\partial\nu|=O(\log y_i)$ and $|\partial^2\log f_1(y_i;\bphi)/\partial\nu^2|=O(y_i)$ for large $y_i$.
Thus from expression (\ref{2nd.deriv}), $|f_{1(\phi_k\phi_\ell)}(y_i;\bphi)|$ is bounded above by an quadratic function of $y_i$.
The similar argument shows that $|f_{1(\phi_k\phi_\ell)}(y_i;\bphi)|$ is bounded above by an cubic function of $y_i$.
Hence, conditional (iii) is satisfied for $a=1$, because $\E[y_i^{c}]<\infty$ for all $c>0$ when $y_i$ has the Poisson-gamma model.
The case of $a=2$ can be shown similarly.

\medskip\noindent
{\bf (Binomial-beta model)}.\ \ \ 
Note that $f_1(y_i;\bphi)$ and $f_2(y_i;\bphi)$ have compact supports and the derivatives $f_{a(\phi_j)}(y_i;\bphi)$, $f_{a(\phi_j\phi_\ell)}(y_i;\bphi)$ and $f_{a(\phi_j\phi_\ell\phi_k)}(y_i;\bphi)$ are finite for an interior point $\bphi$.
Then condition (iii) is easy to check.

\vspace{0.4cm}
% A2 ---Proof of Theorem 1--
\noindent
{\bf A2. Proof of Theorem \ref{thm:numderiv}.}\ \ \ \ 
Let us fix $\bphi_0$ as an interior point of $\Phi$.
We here use the notation $C(y_i)$ as a generic function of $y_i$ with $C(y_i)=O_p(1)$, and the notations b$u_{1i}$ and $u_{2i}\in [-1,1]$ as generic constants.
Expanding $f_a(y_i,\bphi_0+z_m\e_j)$ and $f_a(y_i,\bphi_0-z_m\e_j)$ around $\bphi_0$, we get
\begin{align*}
&f_a(y_i,\bphi_0+z_m\e_j)=f_a(y_i,\bphi_0)+f_{a(\phi_j)}(y_i,\bphi_0)z_m+\frac12f_{a(\phi_j\phi_{\ell})}(y_i,\bphi_0+u_{1i}z_m\e_j)z_m^2\\
&f_a(y_i,\bphi_0-z_m\e_j)=f_a(y_i,\bphi_0)-f_{a(\phi_j)}(y_i,\bphi_0)z_m+\frac12f_{a(\phi_j\phi_{\ell})}(y_i,\bphi_0+u_{1i}z_m\e_j)z_m^2,
\end{align*}
so that it follows that 
\begin{align*}
&(2z_m)|f_{a(\phi_j)}^{\ast}(y_i,\bphi_0)-f_{a(\phi_j)}(y_i,\bphi_0)|\\
&=|f_a(y_i,\bphi_0+z_m\e_j)-f_a(y_i,\bphi_0-z_m\e_j)-2z_mf_{a(\phi_j)}(y_i,\bphi_0)|\\
&=\frac12z_m^2|f_{a(\phi_j\phi_{\ell})}(y_i,\bphi_0+u_{1i}z_m\e_j)-f_{a(\phi_j\phi_{\ell})}(y_i,\bphi_0+u_{2i}z_m\e_j)|\leq C(y_i)z_m^2,
\end{align*}
from (iii) of Assumption \ref{as}.
This shows the first part of Theorem \ref{thm:numderiv}.

To show the other parts, we prove that there exist functions $C_a(y_i)=O_p(1)$ for $a=1,2,3$ such that 
\begin{equation}\label{r.deriv}
|r_{i(\phi_j)}(y_i,\bphi_0)|\leq C_1(y_i),\ \  |r_{i(\phi_j\phi_\ell)}(y_i,\bphi_0)|\leq C_2(y_i), \ \ |r_{i(\phi_j\phi_\ell\phi_k)}(y_i,\bphi_0)|\leq C_3(y_i).
\end{equation}
The straightforward calculation shows that 
\begin{align*}
&r_{i(p)}=f_1f_2\{pf_1+(1-p)f_2\}^{-2}, \ \ \ \ r_{i(pp)}=-2f_1f_2\{pf_1+(1-p)f_2\}^{-3}(f_1-f_2),\\
&r_{i(ppp)}=6f_1f_2\{pf_1+(1-p)f_2\}^{-4}(f_1-f_2)^2,
\end{align*}
which are all bounded above by $C(y_i)$ since $f_a/(pf_1+(1-p)f_2)\leq \max(p^{-1}, (1-p)^{-1})$ for $a=1,2$.
Moreover, it is noted that
\begin{align*}
|r_{i(\phi_j)}|=\frac{p(1-p)\left|f_{1(\psi_j)}f_2-f_1f_{2(\psi_j)}\right|}{\left\{pf_1+(1-p)f_2\right\}^2}\leq \frac{p|f_{1(\psi_j)}|+(1-p)|f_{2(\psi_j)}|}{pf_1+(1-p)f_2}\leq C(y_i)
\end{align*}
under (iii) of Assumption \ref{as}.
Similarly, it can be shown that the higher order derivatives $r_{i(\phi_j\phi_\ell)}$, $r_{i(\phi_j\phi_\ell\phi_k)}$, $r_{i(p\phi_\ell)}$, $r_{i(pp\phi_\ell)}$ and $r_{i(p\phi_\ell\phi_k)}$ have the form $h(y_i,\bphi_0)/\{pf_1+(1-p)f_2\}^c$, where $c$ is a positive integer and $h(y_i,\bphi_0)$ is a polynomial of $f_a, f_{a(\phi_j)}$, $f_{a(\bphi_j\phi_k)}$ and $f_{a(\phi_j\phi_\ell\phi_k)}$, so that there exists $h^{\dagger}(y_i)=O_p(1)$ such that $h(y_i,\bphi_0)\leq h^{\dagger}(y_i)$.
This establishes property (\ref{r.deriv}).
Using the property, we have
\begin{align*}
(2z_m)&\big|\mut_{i(\phi_j)}^{\ast}(y_i,\bphi_0)-\mut_{i(\phi_j)}(y_i,\bphi_0)\big|\\
&=\frac12z_m^2\big|\mut_{i(\phi_j\phi_j)}(y_i,\bphi_0+u_{1i}z_m\e_j)-\mut_{i(\phi_j\phi_j)}(y_i,\bphi_0+u_{2i}z_m\e_j)\leq C(y_i)z_m^2
\end{align*}
and
\begin{align*}
(2z_m)&\big|R_{1i(\phi_j)}^{\ast}(y_i,\bphi_0)-R_{1i(\phi_j)}(y_i,\bphi_0)\big|\\
&=\frac12z_m^2\big|R_{1i(\phi_j\phi_j)}(y_i,\bphi_0+u_{1i}z_m\e_j)-R_{1i(\phi_j\phi_j)}(y_i,\bphi_0+u_{2i}z_m\e_j)\big|\leq C(y_i)z_m^2.
\end{align*}
Finally, we consider the approximation of the second-order partial derivatives of $R_{1i}$.
Expanding $R_{1i}(\bphi_0+z_m\e_j)$ and $R_{1i}(\bphi_0-z_m\e_j)$ up to $O(z_m^3)$, we have
\begin{align*}
z_m^2&\big|R_{1i(\phi_j\phi_j)}^{\ast}(y_i,\bphi_0)-R_{1i(\phi_j\phi_j)}(y_i,\bphi_0)\big|\\
&=\frac16z_m^3\big|R_{1i(\phi_j\phi_j\phi_j)}(y_i,\bphi_0+u_{1i}z_m\e_j)-R_{1i(\phi_j\phi_j\phi_j)}(y_i,\bphi_0+u_{2i}z_m\e_j)\big|\leq C(y_i)z_m^3,
\end{align*}
from property (\ref{r.deriv}).
From this result, we obtain for $j\neq \ell$,
\begin{align*}
R_{1i(\phi_j\phi_\ell)}^{\ast}(y_i,\bphi_0)=\frac1{2z_m^2}&\Big[\left\{R_{1i}(y_i,\bphi_0+z_m\e_{j\ell})+R_{1i}(y_i,\bphi_0-z_m\e_{j\ell})-2R_{1i}(y_i,\bphi_0)\right\}\\
&\hspace{1cm} -z_m^2\left\{R_{1i(\phi_j\phi_j)}(y_i,\bphi_0)+R_{1i(\phi_\ell\phi_\ell)}(y_i,\bphi_0)\right\}\Big]+O_p(z_m).
\end{align*}
It is noted that
\begin{align*}
R_{1i}(y_i,&\bphi_0+z_m\e_{j\ell})=R_{1i}(y_i,\bphi_0)+R_{1i(\psi_j)}(y_i,\bphi_0)z_m+R_{1i(\phi_\ell)}(y_i,\bphi_0)z_m+R_{1i(\phi_j\phi_\ell)}(y_i,\bphi_0)z_m^2\\
&+\frac12R_{1i(\phi_j\phi_j)}(y_i,\bphi_0)z_m^2+\frac12R_{1i(\phi_\ell\phi_\ell)}(y_i,\bphi_0)z_m^2+\frac16z_m^3\sum_{j,\ell,k}R_{1i(\phi_j\phi_\ell\phi_k)}(y_i,\bphi_0+z_mu_{1j\ell k}\e_{j\ell k})
\end{align*}
where $u_{1j\ell k}\in [-1,1]$ and $\e_{j\ell k}=\e_j+\e_{\ell}+\e_k$.
Then it follows that
\begin{align*}
z_m^2&|R_{1i(\phi_j\phi_\ell)}^{\ast}(y_i,\bphi_0)-R_{1i(\phi_j\phi_\ell)}(y_i,\bphi_0)\big|\\
&=\frac16z_m^3\bigg|\sum_{j,\ell,k}R_{1i(\phi_j\phi_\ell\phi_k)}(y_i,\bphi_0+z_mu_{1j\ell k}\e_{j\ell k})-\sum_{j,\ell,k}R_{1i(\phi_j\phi_\ell\phi_k)}(y_i,\bphi_0+z_mu_{2j\ell k}\e_{j\ell k})\bigg|,
\end{align*}
for some $u_{2j\ell k}\in [-1,1]$.
Using property of (\ref{r.deriv}), we conclude that the above term is bounded above by $C(y_i)z_m^3$, which completes the proof.

%-------------------------------------------------------------------------------------------------------------------------------
%    REFERENCE
%-------------------------------------------------------------------------------------------------------------------------------

\end{document}